\documentclass[reprint,amsmath,amssymb,aps,prb,showpacs,floatfix,superscriptaddress]{revtex4-1}

\usepackage{graphicx}
\usepackage{bm}
\usepackage{multirow}
\usepackage{verbatim}
\usepackage{longtable}
\usepackage{color}
\usepackage{nccmath}

\usepackage[normalem]{ulem}

\renewcommand{\figurename}{Fig.}

\newcommand{\sectionname}{Section}

\newcommand{\e}{e}
\renewcommand{\i}{i}
\newcommand{\mc}{\mathcal}
\newcommand{\mr}{\mathrm}

\newcommand{\vphi}{\varphi}
\newcommand{\pd}{\partial}
\newcommand{\dif}{\mathrm{d}}

\newcommand\ph[1]{\phantom{#1}}
\newcommand\abs[1]{\lvert#1\rvert}

\newcommand\avgs[1]{\langle#1\rangle}

\newcommand\bra[1]{\langle#1\rvert}
\newcommand\ket[1]{\lvert#1\rangle}

\newcommand{\Imag}{\operatorname{Im}}

\newcommand{\up}{\uparrow}
\newcommand{\down}{\downarrow}

\newcommand{\cd}{c^{\dag}}
\newcommand{\can}{c^{\phantom{\dag}}}

\newcommand{\ad}{d^{\dag}}
\newcommand{\aan}{d^{\phantom{\dag}}}
\newcommand{\dd}{d^{\dag}}

\newcommand{\hatt}{}
\newcommand{\shs}[1]{}

\begin{document}

\title{Transport in serial spinful multiple-dot systems: \\ The role of electron-electron interactions and coherences}

\author{Bahareh Goldozian}
\affiliation{Mathematical Physics and NanoLund, Lund University, Box 118, S-22100 Lund, Sweden}
\author{Fikeraddis A. Damtie}
\affiliation{Mathematical Physics and NanoLund, Lund University, Box 118, S-22100 Lund, Sweden}
\author{Gediminas Kir{\v{s}}anskas}
\affiliation{Mathematical Physics and NanoLund, Lund University, Box 118, S-22100 Lund, Sweden}
\author{Andreas Wacker}
\affiliation{Mathematical Physics and NanoLund, Lund University, Box 118, S-22100 Lund, Sweden}

\date{\today}

\begin{abstract}
Quantum dots are nanoscopic systems, where carriers are confined in all three spatial directions. Such nanoscopic systems are suitable for fundamental studies of quantum mechanics and are candidates for applications such as quantum information processing. It was also proposed that linear arrangements of quantum dots could be used as quantum cascade laser. In this work we study the impact of electron-electron interactions on transport in a spinful serial triple quantum dot system weakly coupled to two leads. We find that due to electron-electron scattering processes the transport is enabled beyond the common single-particle transmission channels. This shows that the scenario in the serial quantum dots intrinsically deviates from layered structures such as quantum cascade lasers, where the presence of well-defined single-particle resonances between neighboring levels are crucial for device operation.  Additionally, we check the validity of the Pauli master equation by comparing it with the first-order von Neumann approach. Here we demonstrate that coherences are of relevance if the energy spacing of the eigenstates is smaller than the lead transition rate multiplied by $\hbar$.
\end{abstract}

\maketitle

\section{Introduction}

Electron-electron interaction effects in quantum dots have been a topic of active research within the past decades.~\cite{DevoretNature1992,TaruchaPRL1996,KouwenhovenScience1997,ReimannRMP2002,KorkusinskiPRB2007} The ability to confine a finite number of charged particles with advanced fabrication techniques in these structures opened up a possibility for testing different physical theories such as charge and conductance quantization,~\cite{KastnerRMP1992,AverinPRB1991,FultonPRL1987,AshooriNature1996,KouwenhovenZPhysBCondMat1991} Coulomb blockade,~\cite{AverinJLTP1986,YoffeAdvancesinPhysics2001} exciton formation,~\cite{BeardPCL2011,KlimovARPC2007} just to mention a few.

Here we focus on the electric transport through a serial arrangement of multiple quantum dots. Experimentally they can be realized in different ways and prominent examples are the gating of a two-dimensional electron gas,~\cite{WaughPRL1995,FujisawaScience1998,LiuNatPhot2010} cleaved edge overgrowth structures,~\cite{SchedelbeckScience1997} stacked self-organized quantum dots,~\cite{BorgstromAPL2001} nanowires either with external gates~\cite{MasonScience2004,FasthNL2005} or embedded heterostructures,~\cite{FuhrerNL2007} or the  arrangement of atoms by a scanning tunneling microscope.~\cite{FolschNatNanotechnol2014} The restriction of phase space in such low-dimensional structures is  reducing the scattering rates substantially and therefore such structures have been suggested for a wide range of applications ranging from quantum information processing~\cite{LossPRA1998,LairdPRB2010} to quantum cascade lasers.~\cite{BurnettPRB2014,GrangeAPL2014}

Electron transport through these structures has been widely used for level spectroscopy.~\cite{VanDerWielRMP2003,HansonRMP2007,JespersenNatPhys2011,HatanoPRB2013}
Generally, one assumes, that the transport through quantum dot systems is dominated by specific resonances. These occur due to the alignment of energy levels in individual dots with those in neighboring dots as well as with the chemical potentials of metallic leads. This provides specific conditions for transport, which are resolved as current or conductance peaks for varying external parameters, such as the voltages at different gates. Such resonances may even refer to states with different energies due to the emission of phonons with a specified frequency~\cite{BrandesPRL1999,WeberPRL2010} or Auger processes.~\cite{PedersenPRB2007} But even in this case the existence of specific resonances is the guiding theme of studying multiple dot systems. However, with increasing number of dots the number of resonance conditions becomes large and difficult to satisfy simultaneously. Taking into account growth imperfections as well as undefined locations of impurities with fluctuating charges, a strong suppression of current is expected with an increasing number of dots.~\cite{WeymannPRB2011,EmaryPRB2007}

\begin{figure}
\begin{center}
\includegraphics[width=0.95\columnwidth]{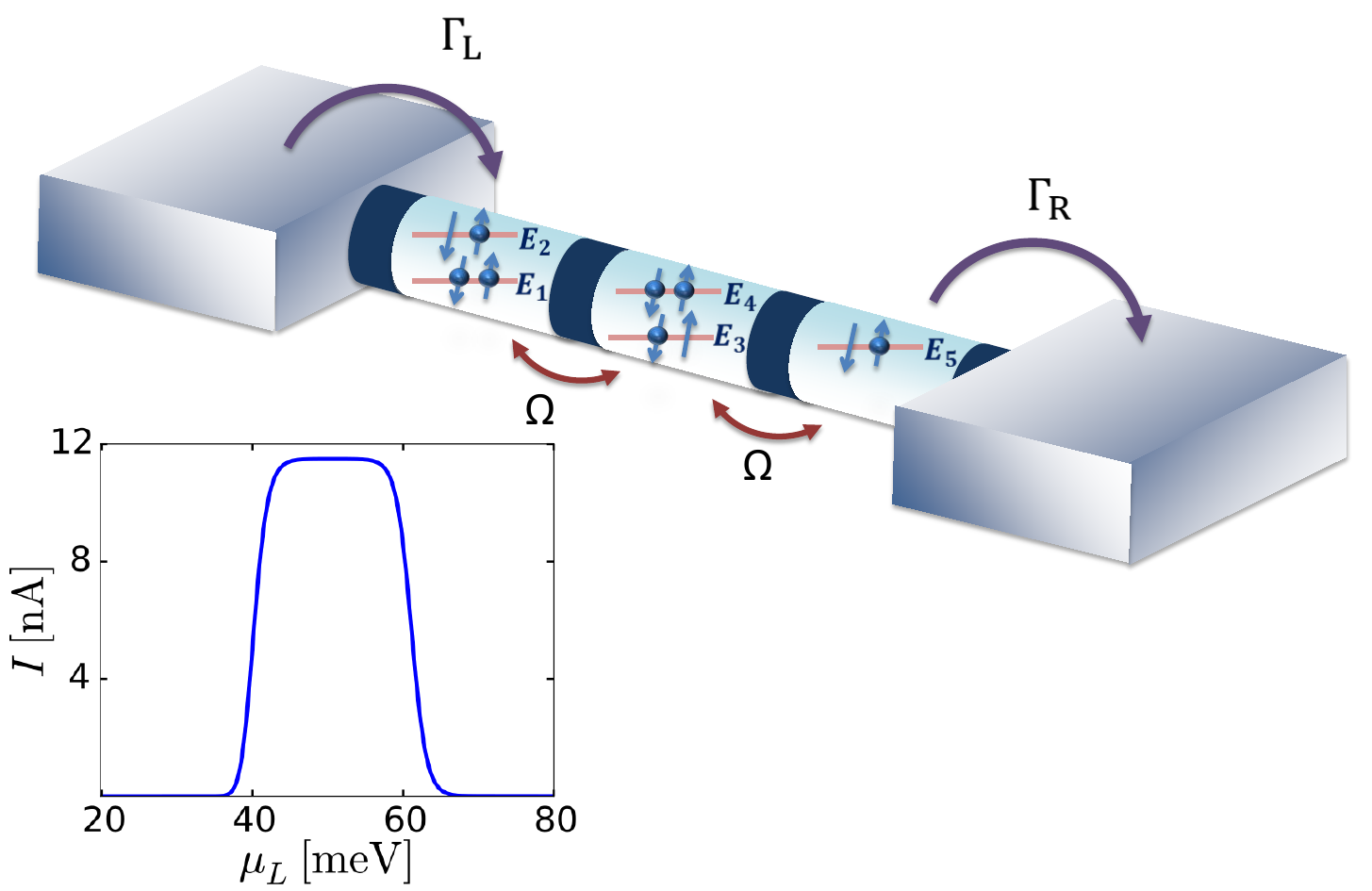}
\caption{\label{fig1} Sketch of the serial triple dot structure attached to the source and drain leads. The electrons from the source/drain lead can tunnel into the left/right dot with the rate $\Gamma_{L}/\Gamma_{R}$. The levels within the dots are coupled by the tunneling matrix $\Omega$.
Inset: Current as a function of the left Fermi level taking into account only the Coulomb scattering term $U_{sc}$. Parameters as in Table~\ref{TableParameters}.}
\end{center}
\end{figure}

Electron-electron interaction is naturally occurring in all electronic devices and affects transport both by scattering (such as the Auger term) and level shifts. For systems with many degrees of freedom, such as  bulk or layered systems, the continuum of states justifies usually a mean-field description, so that one can apply effective single-particle levels with renormalized energies. In this case the resonances occur for different parameters, but the essential principle remains. A very successful example for this concept are quantum cascade lasers, whose operation is based on a clever design for the alignment of such single-particle levels.~\cite{FaistBook2013} For quantum-dot systems, however, any mean-field model is very questionable, as one replaces the interaction between quantized charges  by the interaction of a charge with an averaged quantity.

This work analyzes finite bias electronic transport in both spinless and spinful triple quantum dots coupled in a serial configuration. The considered system serves as a simple model for longer arrangements such as quantum dot superlattices~\cite{GrangePRB2014} or possible dot-based quantum cascade lasers (QCLs).~\cite{BurnettPRB2014,GrangeAPL2014} The main interest of the present work is to address the influence of different parts of the electron-electron ($ee$) interaction inside the triple dot on the electrical current.
Triple quantum dots in other kind of arrangements like triangular shape was extensively studied both theoretically and experimentally.~\cite{SchroerPRB2007,GroveNL2008,WeymannPRB2011,KuzmenkoPRL2006,TrochaPRB2008,EmaryPRB2007,RoggePRB2008}
We show that the Coulomb interaction between electrons opens up a large variety of different channels, which go far beyond the simple pictures of a few resonances, especially when the spin degeneracy of the levels is included. We identify two main  causes: (i) Coulomb scattering provides a generic possibility for energy relaxation within the dot system. (ii) The multitude of many- particle states provides an enormous amount of different possible current paths. Additionally, this multitude of possible current paths give the possibility to study the applicability of simple Pauli master equation approach in realistic physical setup. In this work, we contrast the transport results obtained by Pauli master equation and
first-order von Neumann approach.~\cite{BreuerBook2006,PedersenPRB2007}

\section{Methods}

The system under consideration consists of three serial quantum dots sandwiched between metallic leads (see Fig.~\ref{fig1}). In this section we specify the Hamiltonian, the different approaches for transport, and the parameters used in our calculations.

\subsection{Hamiltonian}
The serial triple dot is modelled by the following Hamiltonian:
\begin{equation}\label{Hamiltonian}
\hatt{H}=\hatt{H}_{\mr{LR}}+\hatt{H}_{\mr{T}}+\hatt{H}_{\mr{D}},
\end{equation}
where
\begin{equation}\label{hamLR}
\hatt{H}_{\mr{LR}}=\sum_{k\ell\sigma}E_{k\ell}^{\ph{}}\cd_{k\ell\sigma}\can_{k\ell\sigma},\\
\end{equation}
describes the source and drain leads as reservoirs with  a continuum of noninteracting electrons, where $\cd_{k\ell\sigma}$ denote the electron creation operators in the leads. Here $\ell=L,R$ stands for the left or right lead, $\sigma=\up,\down$ denotes the spin of the electron, and $E_{k\ell}=E_{k}+\mu_{\ell}$ is the single particle energy of the electron in state $k$. The dispersion $E_{k}$ in the lead is shifted by the chemical potential $\mu_{\ell}$ of the respective lead $\ell$.
Also it is assumed that the lead states constitute a continuum with $E_{k}\in[-D,D]$ having a bandwidth $2D$ and a constant density of states $\nu_{F}$ at the Fermi level.
The dots are coupled to the leads by the tunneling Hamiltonian:
\begin{equation}\label{hamT}
\hatt{H}_{\mr{T}}=\sum_{n,k\ell\sigma}\left(t_{n\ell}\ad_{n\sigma}\can_{k\ell\sigma}+\mathrm{h.c.}\right),
\end{equation}
with $\dd_{k\ell\sigma}$ being electron creation operator in the dots, where $n\in\{1,2,3,4,5\}$ labels single-particle dot states as depicted in \figurename~\ref{fig1}. Only the levels of the left ($n=1,2$) and right ($n=5$) dots are directly coupled to the left and right lead, respectively. Additionally, we assume that all couplings to the leads have the same magnitude and phase. This means that we have the following non-vanishing tunneling amplitudes
\begin{equation}
t_{1L}=t_{2L}=t_{L}, \quad t_{5R}=t_{R},
\end{equation}
which are parameterized by the tunneling rates $\frac{1}{\hbar}\Gamma_{L/R}=\frac{2\pi}{\hbar}\nu_{F}\abs{t_{L/R}}^2$.
We note that for symmetric structures a sixth level in the third dot below $E_5$ is expected. However, it is not considered here, as it hardly contributes to the current flow.

The Hamiltonian of the dots is given by
\begin{equation}\label{hamD}
\hatt{H}_{\mr{D}}=\sum_{n\sigma}E_{n}\ad_{n\sigma}\aan_{n\sigma}
+\sum_{nm\sigma}\Omega_{nm}\ad_{n\sigma}\aan_{m \sigma}
+\hatt{H}_{ee},
\end{equation}
where $\Omega_{nm}$ describes  the coupling between states $n$ and $m$ in neighboring dots. The electron-electron ($ee$) interaction is described by the Coulomb Hamiltonian
\begin{equation}\label{e-eHamiltonian}
\hatt{H}_{ee}=\frac{1}{2}\sum_{\substack{mnkl \\ \sigma\sigma'}}
V_{mnkl}a_{m\sigma}^{\dagger}a_{n\sigma'}^{\dagger}a_{k\sigma'}a_{l\sigma}.
\end{equation}
For a system that has more than one confined electron, this part plays an important role and we will discuss the different matrix elements $V_{mnkl}$ in the following subsection. In general the Coulomb matrix elements read
\begin{equation}\label{V(mnkl)}
V_{mnkl}=\frac{e^2}{4\pi\varepsilon_{r}\varepsilon_{0}}
\int \dif^{3}r\int\dif^{3}r'\frac{\varphi_{m}^{\ast}(\mathbf{r})
\varphi_{n}^{\ast}(\mathbf{r}')\varphi_{k}(\mathbf{r}')\varphi_{l}(\mathbf{r})}
{|\mathbf{r}-\mathbf{r}'|},
\end{equation}
where $\varphi_{m}^{\ast}(\mathbf{r})$ is the spatial part of the single particle state $m$, $\varepsilon_{r}$ and $\varepsilon_{0}$ are the relative and vacuum permittivity, respectively. We neglect all terms, where either  $m$ and $l$ or $k$ and $n$ belong to different quantum dots, as their overlap would be vanishingly small. Furthermore, terms connecting levels of next-nearest neighboring dots are small and neglected as well. The remaining terms can be categorized into \textit{Intradot} and \textit{Interdot} interactions and are separately treated below.

\subsubsection{Intradot Interaction}
For intradot interaction all the levels $mnkl$ are considered to be in the same dot. By employing the normalization condition for the wave function, the direct elements can be estimated as:
 \begin{equation}\label{Intradot}
V_{mnnm}\approx\frac{e^{2}}{4\pi\varepsilon_{r}\varepsilon_{0}\sigma}=U,
\end{equation}
where $\sigma=\sqrt{\langle ({\bf r}-\langle{\bf r}\rangle)^2\rangle}$ is the standard deviation for the spatial extension of the dot wave functions. Another set of interaction matrix elements that has to be taken into account are $V_{mnmn}$ (for $n \neq m$), which act as exchange terms for equal spins and scattering terms for different spins. Trying different test wave functions, we observe
\begin{equation}
V_{mnmn}\approx U_{ex}  \quad \text{with} \quad U_{ex}=\frac{U}{5}.
\end{equation}
\subsubsection{Interdot Interaction}

The direct interaction between two states in the neighboring dot can be approximated in a similar way as Eq.~\eqref{Intradot}:
\begin{equation}\label{Interdot}
V_{mnnm}\approx\frac{e^{2}}{4\pi\varepsilon_{r}\varepsilon_{0}d}=U_n,
\end{equation}
where $d$ is the distance between the centers of the dots. The terms with different combinations of indices, are estimated by a Taylor expansion of $1/|\mathbf{r}-{\mathbf{r}}'|$ around the centers of the respective dots $\mathbf{R}_i,\mathbf{R}_j$, see Ref.~[\onlinecite{ParsonBook2007}]. Using $\abs{\mathbf{R}_i-\mathbf{R}_j}=d$ for neighboring dots, we find
\begin{align}
\label{dc}
V_{lnml}&\approx\frac{e^{2}}{4\pi\varepsilon_{r}\varepsilon_{0}}\frac{\mathbf{s}_{nm}\cdot(\mathbf{R}_i-\mathbf{R}_j)}{d^{3}}=\pm U_{dc},\\
\label{Sc}
V_{mnkl}&\approx\frac{-e^{2}}{4\pi\varepsilon_{r}\varepsilon_{0}}\frac{2\mathbf{s}_{ml}\cdot\mathbf{s}_{nk}}{d^{3}}=U_{sc},
\end{align}
with the intradot dipole matrix element
\begin{equation}\label{dipolematrix}
\mathbf{s}_{nm}=\int \dif^3r\varphi_{n}^{\ast}(\mathbf{r})\mathbf{r}\varphi_{m}(\mathbf{r}).
\end{equation}
The $U_{dc}$ and $U_{sc}$ terms can be interpreted as a dipole-charge interaction and dipole-dipole scattering terms, respectively. The $U_{sc}$ term is responsible for the Auger process and is crucial for the current flow in our system. The sign of $U_{dc}$ in Eq.~\eqref{dc}, depends on whether the charge is on the right or left side of the dipole.
\subsection{Parameter values}
In order to obtain realistic and consistent parameters, we consider a particular nanostructure which is made from a nanowire containing three InAs wells (thickness 40 nm) embedded between InP barriers (thickness 3 nm). Similar structures have been recently fabricated.~\cite{FuhrerNL2007,LindgrenNT2013,JurgilaitisStructuralDynamics2014} The values of the energies $E_i$ and couplings $\Omega_{nm}$ are estimated by a tight-binding superlattice model (see Ref.~[\onlinecite{WackerPR2002}]), as outlined in the \appendixname~\ref{sec:Parameters}. This model also provides the dipole matrix element $|\mathbf{s}_{21}|=8 \ \mathrm{nm}$, which points in the direction of the nanowire, and is used  to estimate the Coulomb matrix elements as outlined above. Furthermore we use $\sigma=11$ nm and $d=43$ nm as well as $\varepsilon_{r,\textrm{InAs}}=15$.~\cite{JoyceNanotechnology2013} In addition, the Coulomb matrix elements were calculated numerically from Eq.~(\ref{V(mnkl)}) using wave functions in cuboids (40 nm $\times$ 35 nm $\times$ 33 nm) representing each dot. This result was in good agreement (about 10\% deviations) with the approximations addressed above. As we are only interested in general features, rather than in modelling a specific device, we used rounded values for all quantities in order to allow for an easy recognition of scales in the plots. The specific values are given in Table~\ref{TableParameters}.

\begin{table}
\begin{center}
\begin{tabular}{c|c|c|c}
  \hline
         $E_{1}=40$ & $U=10$        & $\Gamma_{L}=0.1$ & $\Omega_{13}=-0.05$ \\ 
         $E_{2}=60$ & $U_{ex}=2$    & $\Gamma_{R}=0.1$ & $\Omega_{14}=0.1$ \\ 
         $E_{3}=20$ & $U_{n}=3$     & $\mu_{L}=50$     & $\Omega_{23}=0.1$ \\ 
         $E_{4}=40$ & $U_{sc}=-0.2$ & $\mu_{R}=10$     & $\Omega_{24}=0.2$ \\ 
         $E_{5}=20$ & $U_{dc}=-0.5$ & $k_{\mr{B}}T=1$  & $\Omega_{35}=0.1$ \\ 
                    &               & $D=10^4$           & $\Omega_{45}=0.2$ \\
  \hline
\end{tabular}
\end{center}
\caption{\label{TableParameters} Parameters used in the calculations if not mentioned otherwise. All energies are in meV. }
\end{table}

\subsection{\label{Sec:Method}Transport Calculation}

We are interested in obtaining the current from the left to the right lead for considerably large bias. In order to do that, we diagonalize the Hamiltonian $\hatt{H}_{\mr{D}}$~\eqref{hamD} and get the many-particle eigenstates $ |a\rangle,|b\rangle,\ldots$ of the triple dot. Expressed in this many-particle basis the tunneling Hamiltonian $\hatt{H}_{\mr{T}}$~\eqref{hamT} becomes
\begin{align}
\label{hamT2}
&\hatt{H}_{\mr{T}}=\sum_{ab,k\ell\sigma}\left(T_{ba}(\ell\sigma)\ket{b}\bra{a}\can_{k\ell\sigma}+\mathrm{h.c.}\right),\\
&T_{ba}(\ell\sigma)=\sum_{n}t_{n\ell}\bra{b}\ad_{n\sigma}\ket{a}.
\end{align}
Here we used the \textit{letter convention}: if more than one state enters an equation, then the position of the letter in the alphabet follows the particle number (for example $N_b= N_a +1$, $N_c= N_a +2$, $N_{a'}=N_a$). In such a way the sum $\sum_{bc}$ restricts to those combinations, where $N_c = N_b+1$. To obtain the current through the device we use the first-order von Neumann (1vN) approach,~\cite{PedersenPRB2007} where coherent effects are included, and a simpler Pauli master equation where only populations are considered.~\cite{KinaretPRB1992,PfannkuchePRL1995,CavaliereNJP2009} The derivation of the governing equations is presented in \appendixname~\ref{sec:1vN} and \appendixname~\ref{sec:Pauli}.
The 1vN approach treats the reduced density matrix in lowest order with respect to $\hatt{H}_{\mr{T}}$. It is conceptually similar to the Wangsness-Bloch-Redfield~\cite{BreuerBook2006} (also presented in \appendixname~\ref{sec:1vN}), albeit, the Markov limit is done in a different way. For our calculations we did not observe any differences in the current between these approaches as long as principal part integrals are neglected. We choose the temperature of the leads to be larger than the lead tunneling rate, $k_BT\gg\Gamma$, which justifies the neglect of higher-order tunneling processes in the 1vN and the Wangsness-Bloch-Redfield approach. Additionally, this suppresses any kind of Kondo correlations (dominating below the Kondo temperature $T_{K}$), as  $k_BT_K < \Gamma$.~\cite{HewsonBook1993}

Note that these approaches take into account electron-electron scattering by using the eigenstates of the dot-Hamiltonian. The Auger term of the Coulomb interaction with matrix element $U_{sc}$ couples different configurations of many-particle states, which establishes connections between the leads. This does not require any relaxation terms inside the quantum-dot structure. Further relaxation processes, e.~g., due to phonon-scattering, are entirely neglected here. If the emission of optical phonons is energetically not allowed, phonon scattering rates between quantum dot states are of the order of 1/ns or even smaller.~\cite{BockelmannPRB1990,AgamPRL1997,NakaokaPRB2006} Such a scattering process can provide background currents of at most $\approx  0.1$ nA. As this is smaller than the current peaks addressed in the subsequent section, it is justified to neglect these processes in our study.

\section{Results}

\begin{figure*}
\begin{center}
\includegraphics[width=0.95\textwidth]{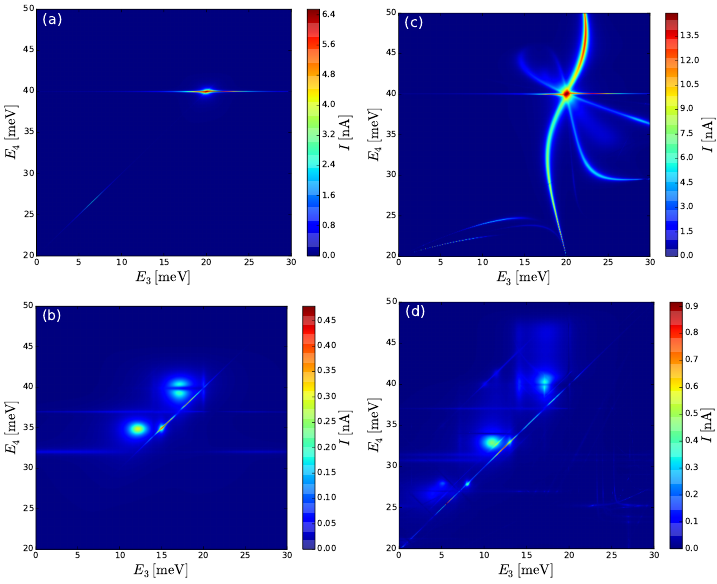}
\caption{\label{fig2} Current through the triple dot system calculated using the Pauli master equation approach. (a) Spin-polarized system with just scattering $U_{\mathrm{sc}}$ included [Eq.~\eqref{Sc}]. (b) Spin-polarized system with all Coulomb matrix elements included. (c) Spinful system with just scattering $U_{\mathrm{sc}}$. (d) Spinful system with all Coulomb matrix elements. Parameters as in Table~\ref{TableParameters}.}
\end{center}
\end{figure*}

\begin{figure*}
\begin{center}
\includegraphics[width=0.95\textwidth]{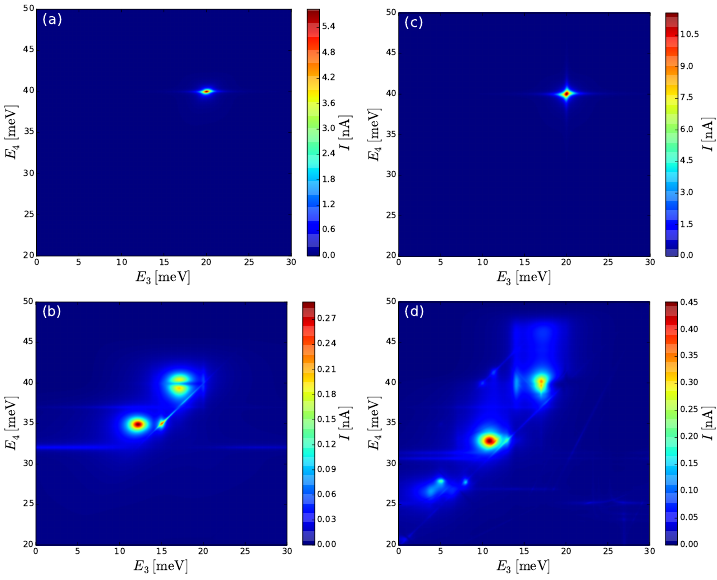}
\caption{\label{fig3} Same as \figurename~\ref{fig2} but with the current calculated using the 1vN approach. (a) The current has a single distinct peak, where $E_{3}\approx E_{5}$ and $E_{4}\approx E_{1}$ and the resonance extends along $E_{4}=40 \ \mathrm{meV}$ line. (b) Inclusion of all possible Coulomb elements opens up additional channels for transport leading to more current peaks. (c) The maximum current becomes larger by a factor of two compared to the spin-polarized case, however, now the resonance extends faintly along the $E_{4}=40 \ \mathrm{meV}$ \textit{and} $E_{3}=20 \ \mathrm{meV}$ lines. (d) For spinful system additional Coulomb elements open very many channels, which leads to \textit{significant} background current. Parameters as in Table~\ref{TableParameters}. }
\end{center}
\end{figure*}
\begin{figure}
\begin{center}
\includegraphics[width=0.99\columnwidth]{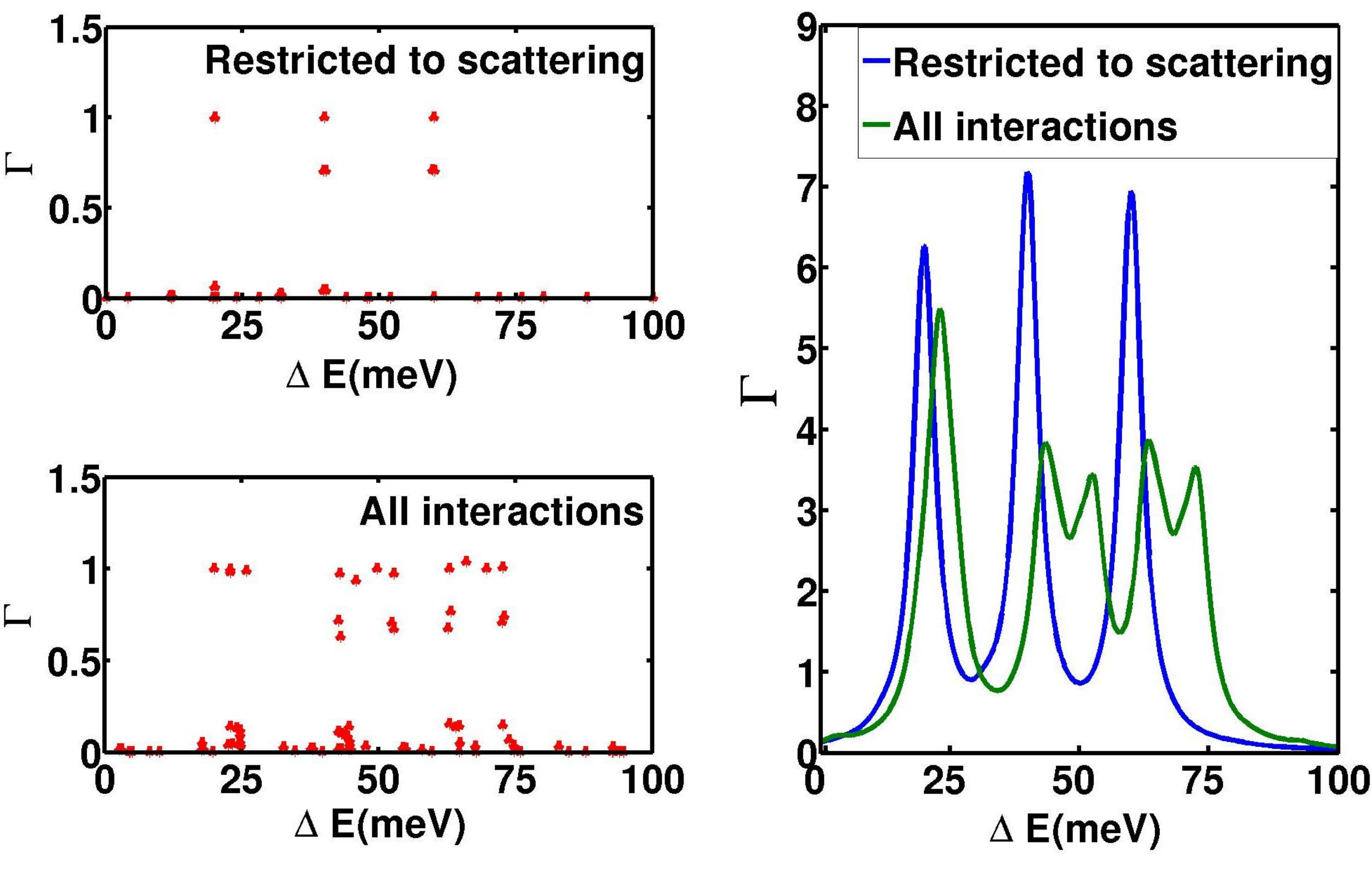}
\caption{\label{fig4} Addition energies (horizontal) for the two-particle states in the spin-polarized system. The vertical axis shows the respective coupling strength for electronic transitions from either lead. In the left panels a point is drawn for each transition. In order to resolve multiple transitions, the right panel sums Lorentzians with full width at half maximum of 5 meV  with a peak given by the points in the left panel.}
\end{center}
\end{figure}
\begin{figure}[!ht]
\begin{center}
\includegraphics[width=0.9\columnwidth]{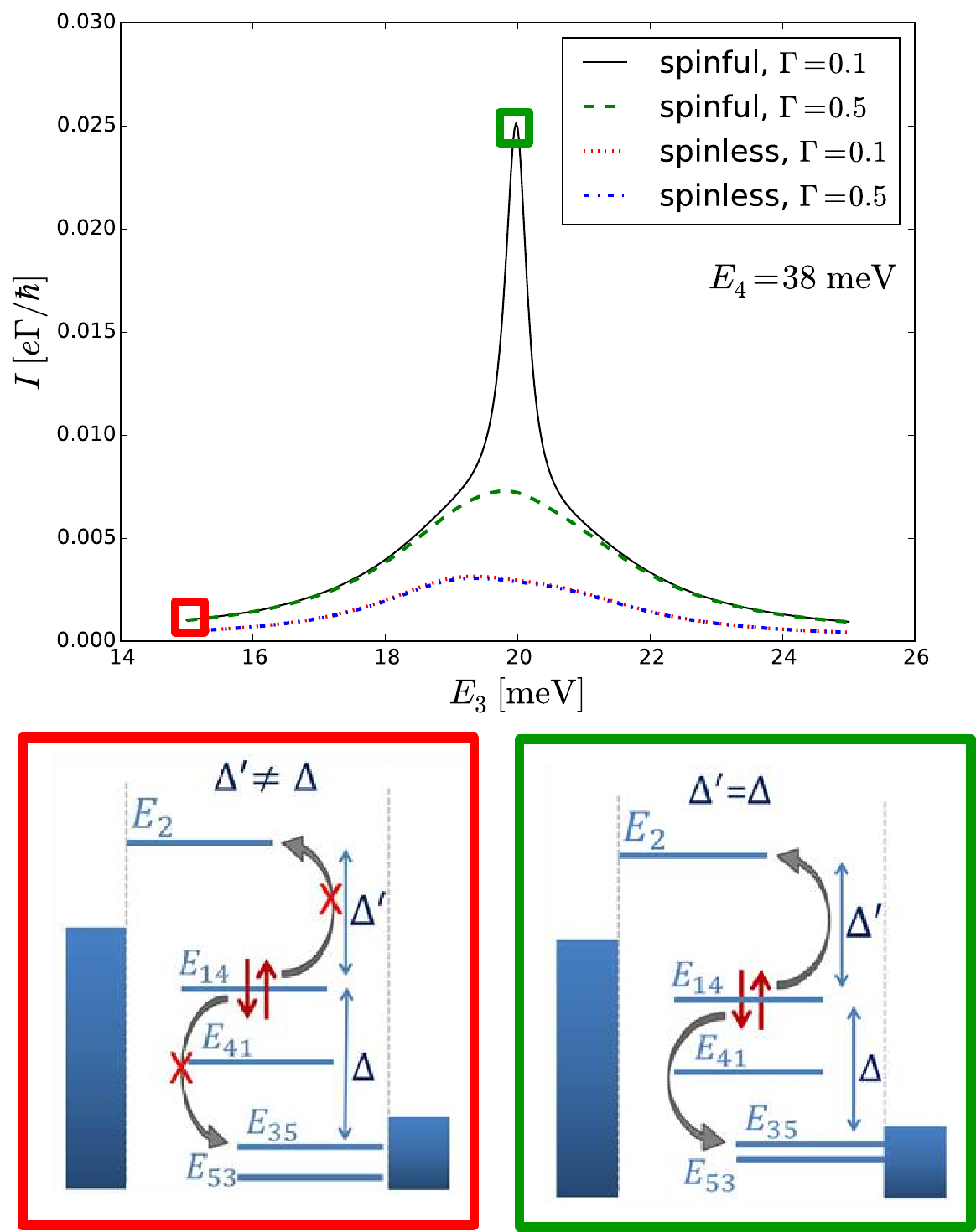}
\caption{\label{fig5} Cuts of contour plots in \figurename~\ref{fig3}c [solid (black)] and \figurename~\ref{fig3}a [dotted (red)] showing increased current for spinful system due to Auger process. Dashed (green) and dashed dotted (blue) gives current when the coupling $\Gamma$ is increased. The left and right square panels show energy spectrum for the one-particle states at two values of energy $E_3$. The units of $\Gamma$ in the legend are $\mathrm{meV}$. }
\end{center}
\end{figure}

In the following we show calculations for varying levels $E_3$ and $E_4$. This is motivated as follows: even in a nominally identical triple-dot system, growth imperfections, charged impurities and the electrostatic environment will modify the energy levels in a way which is difficult to predict. We fix the excitation energy in the left dot, $E_2-E_1$, as a reference point and the level $E_5$ of the right dot can be shifted by a source drain bias. However, the remaining levels $E_3$ and $E_4$ are less controllable, albeit there is a possibility of gating. Thus it is a central question of practical interest, for which range of parameters $E_3,E_4$ current flow occurs.

For an applied bias, standard considerations of coherent transport, such as the transmission  formalism, predict current flow, if there is a state connecting both leads. If we neglect the Coulomb interaction at all, there is only a very small current unless three levels align in a row (e.g.~for $E_1\approx E_4\approx E_5$, a case we do not consider). For the situation in Fig.~\ref{fig1} one would thus not expect any elastic transport through the  structure. Neglecting phonon scattering, the current would  thus be zero, even if the energies $E_{3}$ and $E_{4}$ are varied. However, in  Ref.~[\onlinecite{LassenPRB2007}] it was shown, that for this situation, electron-electron  scattering enables a current flow: if the levels 1 and 4 are occupied, an energy-conserving scattering event creates the simultaneous transitions 1$\rightarrow$2 and 4$\rightarrow$3 (for $E_{3}$-$E_{4}$ = $E_{2}$-$E_{1}$), with subsequent tunneling to the right lead via the state 5. The inset of Fig.~\ref{fig1} shows the current as a function of left Fermi level. As it is shown the current is decreasing drastically when the left Fermi level comes close to the excited state of the first dot. Then the level $E_{2}$ becomes occupied, and consequently the Auger scattering process is prevented due to the Pauli blocking. This is a distinct feature of the Auger driven transport in multiple dots, which discriminates it from other possible scattering processes. Thus, scattering-assisted processes dominate.

A finite Coulomb scattering matrix element $V_{2341}=U_{\mathrm{sc}}$ allows for an Auger process, where one electron relaxes from level 4 to level 3, while transferring its energy to an electron being excited from level 1 to level 2. Therefore $U_\mathrm{sc}=-0.2$ meV is used in all calculations, while the impact of the other Coulomb terms is neglected in some calculations in order to demonstrate their relevance. The left and right Fermi levels are fixed to be $\mu_{L}=50 \ \mathrm{meV}$, $\mu_{R}=10  \ \mathrm{meV}$ so that the leads fill level 1 and empty the levels 2 and 5. The calculations using the Pauli master equation and the 1vN approach are presented in Figs.~\ref{fig2} and ~\ref{fig3}, respectively.

\subsection{\label{SecSingleSpin} Spin-polarized levels}

At first we restrict to spin-polarized levels in the quantum dot. This could be achieved by having a large magnetic field in the system. Figures~\ref{fig2}a and \ref{fig3}a show the current as a function of $E_{3}$ and $E_{4}$, when just scattering elements $U_{\mathrm{sc}}$ are included. Only for specific values of $E_{3}$ and $E_{4}$ the electrons are able to pass through the system. This reflects the conservation of energy both for the electron-electron scattering and for tunneling, i.~e., $E_{3}=E_{5}=20 \ \mathrm{meV}$ and $E_{4}=E_{1}= 40 \ \mathrm{meV}$. Otherwise, the electron transport is blocked (or limited to a background current below
$0.1 \ \mathrm{nA}$ due to phonon-scattering in real structures as discussed above). Thus, any slight changes in the geometry of the dots and the configuration of the energy levels would prevent the current flow. Such a selective situation would be optimal for devices, which rely on well defined transitions, such as quantum cascade lasers.

When $ee$-interaction with all the matrix elements is considered, we see that the current can flow for a wider range of parameters, as shown in Figs.~\ref{fig2}b and \ref{fig3}b. This scenario is reflected by the addition energies, where a particle can tunnel into or out of the triple dot. In Fig.~\ref{fig4} we display the differences in energy between all possible  two and three-particle states ($\Delta{E}$ axis). In order to specify their relevance for single-particle transitions, we plot the respective transition probabilities for electrons to enter from either lead, while changing the state of the system between these states. For the case restricting to Coulomb scattering, there are only three distinct energies, where lead electrons can enter, which correspond to the energies of the levels 1, 2, and 5. In contrast, a larger variety of excitation energies is relevant if the full Coulomb interaction is taken into account. This is fully consistent with the differences between the current plots of Figs.~\ref{fig3}a and Fig.~\ref{fig3}b.

\subsection{Spinful levels}

In the absence of an external magnetic field each single-particle energy level is doubly degenerate due to the Kramer's theorem. Let us first consider the case with just scattering $U_{\mathrm{sc}}$ elements present (Fig.~\ref{fig3}c). In a simple picture, one could expect, that the current doubles compared to the case where the levels are spin-polarized (Fig.~\ref{fig3}a). The maximum of the current in Fig.~\ref{fig3}c shows indeed an increase by a factor of two. Furthermore, Fig.~\ref{fig3}c shows that the resonance faintly extends along the intersection of $E_{3}=20$ and $E_{4}=40$ lines. However, in a spinless case this resonance is extending just along the $E_{4}=40$ line.

In order to understand the reason of the different behavior in \figurename~\ref{fig3}, we study a line for fixed $E_{4}=38 \ \mathrm{meV}$ in \figurename~\ref{fig5} and corresponding eigenstates as indicated by the squares. For parameter values at the green (right) square significant current is observed for the spinful levels, but current is blocked for the spin-polarized levels. At the red (left) square current is blocked for both cases.

The five one-particle eigenstates of the Hamiltonian responsible for transport are illustrated in the green diagram in the lower diagrams of Fig.~\ref{fig5}. There is a strong coupling between $E_{1}$ and $E_{4}$ as well as between $E_{3}$ and $E_{5}$ thus the superposition of each of these two states creates two new states with two new energies, which are referred to as $E_{1,4}$, $E_{4,1}$, $E_{3,5}$, and $E_{5,3}$. For the parameters at the right (green) square the conservation of energy is satisfied, $(E_{2}-E_{1,4}) \approx  (E_{1,4}-E_{3,5})$, which allows for Coulomb scattering. However, the state $E_{1,4}$ appears as the initial state in both parts for an Auger process. Thus, the Pauli principle can only be satisfied if the level is spin degenerate. Due to this reason the electrons are able to transfer through the triple dot in a spinful system, but the current is blocked in a spin-polarized system. Spin is definitely more than a factor of two here.
We note that this transport channel is rather sensitive to actual couplings to the leads. The current is \textit{reduced} if the coupling to the leads is \textit{increased} as can be seen from dashed (green) curve. The coherences are responsible for this surprising reduction as discussed in the next section.

Including all $ee$-interaction terms for spinful system provides a wide variety of single particle excitations and consequently resonance conditions are easier to satisfy than for the spin-polarized case addressed above. Figures~\ref{fig2}d and \ref{fig3}d show the evaluated current and we observe a multitude of peaks as well as \textit{significant} background current of $\sim0.15 \ \mathrm{nA}$ for a large number of energy level combinations. As each peak relates to a different current path through the structure, it is very difficult to address or identify a specific transport path in an experiment.

\subsection{Role of coherences: comparison of 1vN and Pauli approaches}

In the following, we discuss the role of coherences, which in the 1vN approach are defined as the off-diagonal elements of the reduced density matrix in the many-body eigenbasis of $H_{\mr{D}}$ (see \appendixname~\ref{sec:1vN}). The Pauli master equation approach neglects any such coherences, which are known to be of relevance, if $\Delta E \lesssim \Gamma$, where $\Delta E$ is the separation between two relevant levels and $\Gamma$ is the transition rate in units of energy. Typical examples for this situation have been already discussed in Refs.~[\onlinecite{WackerPRL1998,CallebautJAP2005}]. In order to verify our results obtained by the Pauli master equation, we compare the 1vN approach,~\cite{PedersenPRB2007} which keeps the coherences of the system and takes into account the tunnel transitions to the leads in lowest order. If the temperature $k_{\mathrm{B}}T$ of the leads surpasses the transition rate $\Gamma$, the 1vN approach is believed to provide reliable results for the currents, as shown by comparison with higher-order approaches.~\cite{PedersenPRB2007}

By comparing Figs.~\ref{fig2} and \ref{fig3} we see that if all interactions are included [(b) and (d)] the peak structure in the 1vN and Pauli approaches is similar. However, the thin line of resonance for $E_4\approx E_3+20$ is substantially reduced in the more advanced 1vN simulation. Along this line the 2-particle state occupying the levels 1 and 3, and the 2-particle state occupying levels 4 and 5 are degenerate. Via the second-order couplings they get mixed and anticross with a small splitting of $\Delta E\approx 2 \ \mu\mathrm{eV}$.  This energy difference is much smaller than the transition rates $\Gamma=0.1$, which explains, why this narrow line is mostly an artificial result in the Pauli master equation. The splitting of the 2-particle states is enhanced to $\Delta E\approx 0.1$ around $E_4=35$, $E_3=15$ for spinless and $E_{4}=32.5$, $E_{3}=13$ for spinful cases, where further 2-particle states are in resonance. This corresponds to a broader peak, which is also visible in the 1vN simulation. If the coupling to the leads is increased, coherences become even more important, which leads to further reduction of these resonances. 
On the other hand, we note that the 1vN calculations match the Pauli master equation result if the coupling to the leads $\Gamma$ becomes vanishingly small compared to the energy $\Delta{E}$ splittings between the many-particle states with the same number of particles, i.~e. $\Gamma/\Delta{E}\rightarrow 0$.

Similar considerations hold for the case when just scattering $U_{sc}$ is included [see (a) and (c) in Figs.~\ref{fig2} and \ref{fig3}]: for spin-polarized levels (a) thin lines of resonances with high current appear at $E_4\approx 40$ and $E_4\approx E_3+20$ (faintly) for the Pauli master equation, which gets suppressed by coherences. However, in the spinful case (c) the Pauli master equation provides a very different scenario compared to the 1vN approach. In \figurename~\ref{fig2}c  many resonances are pronounced, which get completely diminished by the 1vN approach in \figurename~\ref{fig3}c, leaving just an extended resonance along the intersection of $E_{3}=20$ and $E_{4}=40$ lines.

\subsection{\label{sec:nprinp}Neglect of principal part terms}

\begin{figure}
\begin{center}
\includegraphics[width=0.9\columnwidth]{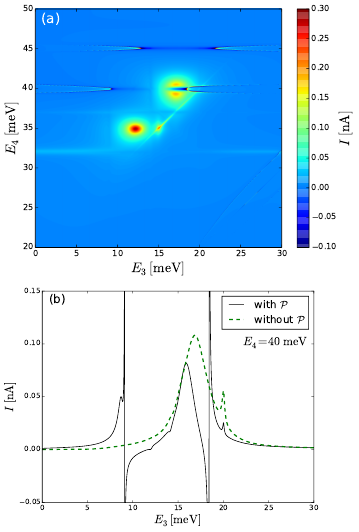}
\caption{\label{fig6} Current for spin-polarized system with all interactions calculated using the 1vN approach [similar to \figurename~\ref{fig3}b], when the principal part $\mc{P}$ integrals are included [see Eq.~\eqref{ooxi}]. (a) Due to divergent current the values below $-0.1 \ \mathrm{nA}$ and above $+0.3 \ \mathrm{nA}$ are filtered. (b) Cut showing the comparison of calculated current with and without the principal parts. }
\end{center}
\end{figure}

In the 1vN approach calculations of Figs.~\ref{fig1}-\ref{fig5} we have In the 1vN approach calculations of Figs.~\ref{fig1}-\ref{fig5} we have neglected the principal part integrals [see Eq.~\eqref{ooxi}]. At the resonance position $p_{ba}^{\ell}=0$ this integral gives a logarithmic divergence, which is cut-off by the temperature. So it could be expected that for large enough temperature these terms should not be problematic. However, in \figurename~\ref{fig6}, we show a calculation for spin-polarized quantum dots, with all interactions and the principal part integrals included. We see that regions of negative current appear (flow against the bias), and at particular points the current also has divergencies (see \figurename~\ref{fig6}b). At these points of divergent current the positivity of the populations $\Phi_{bb}$ is highly violated and the elements $\Phi_{bb'}$ acquire divergent structure as well. This unphysical behavior of the current motivates the neglect of the principal part integrals.

Finally, the principal parts have the structure of the renormalization of the energy spectrum (Lamb shift~\cite{BethePR1947}) as can be seen from perturbation theory for energies~\cite{LowdinJCP1951} or real parts of self-energies in Green's function formalism.~\cite{FetterBook2003} This suggest that this type of terms should be resummed to get an effective Hamiltonian in which new eigenspectrum of the quantum dots is obtained and the energy differences $E_{b}-E_{a}$ are renormalized.~\cite{KirsanskasPRB2012,KollerPRB2012}

\section{Discussion}

The main results are collected in \figurename~\ref{fig3}, which displays the current through both spin-polarized and spinful triple quantum dot system, calculated using the 1vN approach. For the single spin case restricting just to scattering $U_{\mathrm{sc}}$, a single peak is seen in Fig.~\ref{fig3}a, which is easily predicted by standard single particle states. Such resonances are well known and are commonly used for device design.  However, taking into account spin degeneracy of the states and all interactions, \figurename~\ref{fig3}d displays an entirely different picture, where current flow is spread over a wide range of parameters where unexpected paths become of relevance. As a consequence, designing devices based on multiple quantum dots is questionable, if one restricts to single-particle models even if the mean-field is included. For example, for dot-based quantum cascade lasers, it is difficult to accomplish the desired specificity of electron injection into the upper laser level. On the other hand, the multitude of channels creates \textit{significant} background current for wide range of energy level configurations, which makes transport in multiple quantum-dot devices less sensitive to size fluctuations and random charges.

We have compared the Pauli master equation approach, which neglects coherences, and the 1vN approach where all density matrix elements of the reduced quantum dot system are taken into consideration. It was shown that the qualitative picture of resonances is correctly captured by the simpler Pauli master equation. However, there are cases where the current is highly overestimated by this approach and this can be seen by comparing Figs.~\ref{fig2}c and \ref{fig3}c. These points are traced back to specific level configurations with the lead coupling larger than the energy splitting, i.~e., $\Gamma>\Delta{E}$, where coherences strongly reduce the current. This also allows for situations where the current drops with increasing lead coupling $\Gamma$ as shown Fig.~\ref{fig5}. Additionally, the relevance of principal part integrals in the 1vN approach was examined and it was shown that it can lead to highly unphysical results (current flowing against the bias), which suggests that neglecting such terms is reasonable.

\begin{acknowledgments}
This work was supported by the Swedish Research Council (VR), NanoLund.
\end{acknowledgments}

\appendix

\section{\label{sec:1vN}First-order von Neumann (1vN) approach}
The dynamics of the full density matrix $\rho$ of the system is described by the von Neumann equation:
\begin{equation}\label{vneq}
\i\hbar\frac{\pd}{\pd t}\rho=[H,\rho].
\end{equation}
We solve the above equation approximately using the 1vN approach.~\cite{PedersenPRB2005a,PedersenPRB2007,PedersenPHE2010} In this approach only the density matrix elements, which connect the states differing by just one electron or hole excitation, are considered. Additionally, such a treatment is valid if the temperature $T$ is larger than the tunneling rates, $T\gg\Gamma$. In the other limit, $T\ll\Gamma$, different approaches based on numerical renormalization group,~\cite{AndersPRL2008} quantum Monte Carlo simulations,~\cite{HanPRL2007} or higher order expansions should be used.~\cite{PedersenPRB2005a,HershfieldPRL1993,JinJCP2008,SchoellerEurPhysJSpecTop2009} The density matrix elements are defined as
\begin{equation}
\rho_{ag,bg'}^{[n]}=\bra{ag}\rho\ket{bg'},
\end{equation}
where $\ket{bg}=\ket{b}\otimes\ket{g}$, with $\ket{b}$ denoting the eigenstate of the dot Hamiltonian~(5) and $\ket{g}$ denoting the eigenstate of the lead Hamiltonian~(2). Here the label $n$ provides the number of electron or hole excitations needed to transform $\ket{g}$ into $\ket{g'}$. For example, we will consider matrix elements of the type
\begin{equation}
\begin{aligned}
&\rho_{bg,b'g}^{[0]}=\bra{bg}\rho\ket{b'g},\\
&\rho_{bg-\kappa,ag}^{[1]}=\bra{bg-\kappa}\rho\ket{ag}.
\end{aligned}
\end{equation}
Here we have introduced the following notation
\begin{equation}
\kappa\equiv k, \ \ell, \ \sigma;
\end{equation}
\begin{equation}
\begin{aligned}
\ket{bg+\kappa}&=\ket{b}\otimes\cd_{\kappa}\ket{g},\\
\ket{bg-\kappa}&=\ket{b}\otimes\can_{\kappa}\ket{g}.
\end{aligned}
\end{equation}
By neglecting all the density matrix elements with more than one electron or hole excitation $n\geq2$ from Eq.~\eqref{vneq} we obtain the equations
\begin{equation}\label{rheq0}
\begin{aligned}
\i\hbar\frac{\pd}{\pd{t}}\rho_{bg,b'g}^{[0]}&=(E_{b}-E_{b'})\rho_{bg,b'g}^{[0]}\\
&+\shs{\sum_{a_1,\kappa_1}}T_{ba_1}(\kappa_1)\rho_{a_1g+\kappa_1,b'g}^{[1]}(-1)^{N_{a_1}}\\
&+\shs{\sum_{c_1,\kappa_1}}T_{bc_1}(\kappa_1)\rho_{c_1g-\kappa_1,b'g}^{[1]}(-1)^{N_{b}}\\
&-\shs{\sum_{c_1,\kappa_1}}\rho_{bg,c_1g-\kappa_1}^{[1]}(-1)^{N_{b'}}T_{c_1b'}(\kappa_1)\\
&-\shs{\sum_{a_1,\kappa_1}}\rho_{bg,a_1g+\kappa_1}^{[1]}(-1)^{N_{a_1}}T_{a_1b'}(\kappa_1),
\end{aligned}
\end{equation}
\begin{equation}\label{rheq1}
\begin{aligned}
\i\hbar\frac{\pd}{\pd{t}}\rho_{cg-\kappa,bg}^{[1]}&\approx(E_{c}-E_{\kappa}-E_{b})\rho_{cg-\kappa,bg}^{[1]}\\
&+\shs{\sum_{b_1,\kappa_1}}T_{cb_1}(\kappa)\rho_{b_1g,bg}^{[0]}(-1)^{N_{b_1}}\bra{g}\cd_{\kappa}\can_{\kappa}\ket{g}\\
&-\shs{\sum_{c_1,\kappa_1}}\rho_{cg-\kappa,c_1g-\kappa}^{[0]}(-1)^{N_{b}}T_{c_1b}(\kappa).
\end{aligned}
\end{equation}
Here $N_{b}$ denotes the number of electrons in the state $\ket{b}$. Note that all indices with subscript 1 like $a_1, \ c_1, \ \kappa_1$ are summed over and the \textit{letter convention} introduced in the main text is used. Also the matrix element $\bra{g}\cd_{\kappa}\can_{\kappa}\ket{g}$ in the second term of the right-hand side of Eq.~\eqref{rheq1} corresponds to the requirement that there is an electron in the single particle state corresponding to $\kappa$. Additionally, phase factors like $(-1)^{N_{b}}$ appear due to order exchange of the lead operators with the dot operators, i.e., $\can_{\kappa}(\ket{b}\otimes\ket{g})=(-1)^{N_{b}}\ket{b}\otimes\can_{\kappa}\ket{g}$.

Summing Eqs.~\eqref{rheq0} and \eqref{rheq1} over all the lead states $\ket{g}$ we get
\begin{equation}\label{pheq0}
\begin{aligned}
\i\hbar\frac{\pd}{\pd{t}}\Phi_{bb'}^{[0]}&=(E_{b}-E_{b'})\Phi_{bb'}^{[0]}\\
&+\shs{\sum_{a_1,\kappa_1}}T_{ba_1}(\kappa_1)\Phi_{a_1b'}^{[1]}(\kappa_1)
+\shs{\sum_{c_1,\kappa_1}}T_{bc_1}(\kappa_1)\Phi_{c_1b'}^{[1]}(\kappa_1)\\
&-\shs{\sum_{c_1,\kappa_1}}\Phi_{bc_1}^{[1]}(\kappa_1)T_{c_1b'}(\kappa_1)
-\shs{\sum_{a_1,\kappa_1}}\Phi_{ba_1}^{[1]}(\kappa_1)T_{a_1b'}(\kappa_1)
\end{aligned}
\end{equation}
\begin{equation}\label{pheq1}
\begin{aligned}
\i\hbar\frac{\pd}{\pd{t}}\Phi_{cb}^{[1]}(\kappa)&\approx(E_{c}-E_{\kappa}-E_{b})\Phi_{cb}^{[1]}(\kappa)\\
&+\shs{\sum_{b_1,\kappa_1}}T_{cb_1}(\kappa)\Phi_{b_1b}^{[0]}f_{\kappa}
-\shs{\sum_{c_1,,\kappa_1}}\Phi_{cc_1}^{[0]}T_{c_1b}(\kappa)f_{-\kappa},
\end{aligned}
\end{equation}
where we introduced the following notation
\begin{equation}
\begin{aligned}
&\Phi_{bb'}^{[0]}=\sum_{g}\rho_{bg,b'g}^{[0]},\\
&\Phi_{cb}^{[1]}(\kappa)=\sum_{g}\rho_{cg-\kappa,bg}^{[1]}(-1)^{N_{b}},\\
&\Phi_{bc}^{[1]}(\kappa)=\big[\Phi_{cb}^{[1]}(\kappa)\big]^{*},\\
&f_{\kappa}\equiv f_{k\ell}=(\exp[E_{k}/k_{\mr{B}}T_{\ell}]+1)^{-1},\\
&f_{-\kappa}\equiv 1-f_{k\ell}.
\end{aligned}
\end{equation}
Here we have also assumed that the electrons in the leads are \textit{thermally distributed} according to the Fermi-Dirac distribution $f$ and that this distribution is not affected by the coupling to the quantum dots. This assumption leads to the following relations:
\begin{equation}
\begin{aligned}
&\sum_{g}\rho_{bg,b'g}^{[0]}\bra{g}\cd_{\kappa}\can_{\kappa}\ket{g}\approx f_{\kappa}\Phi_{bb'}^{[0]},\\
&\sum_{g}\rho_{bg-\kappa,b'g-\kappa}^{[0]}\approx f_{-\kappa}\Phi_{bb'}^{[0]}.\\
\end{aligned}
\end{equation}

For the stationary state we assume the conditions
\begin{equation}\label{statcond}
\begin{aligned}
&\i\hbar\frac{\pd}{\pd{t}}\Phi_{bb'}^{[0]}=0,\quad
\i\hbar\frac{\pd}{\pd{t}}\Phi_{cb}^{[1]}(\kappa)=0,
\end{aligned}
\end{equation}
which allow to write $\Phi^{[1]}$ in terms of $\Phi^{[0]}$ as
\begin{equation}\label{ssphi1_1vN}
\Phi_{cb}^{[1]}(\kappa)=
\frac{T_{cb_1}(\kappa)\Phi_{b_1b}^{[0]}f_{\kappa}
-\Phi_{cc_1}^{[0]}T_{c_1b}(\kappa)f_{-\kappa}}{E_{\kappa}-E_{c}+E_{b}+\i\eta}.
\end{equation}
Here we have added a positive infinitesimal $\eta=+0$ to ensure a proper decay of initial conditions, which can be seen by formally integrating Eq.~\eqref{pheq1}:
\begin{equation}
\begin{aligned}
&\Phi_{cb}^{[1]}(\kappa,t)=\frac{1}{\i\hbar}\int_{-\infty}^{t}\dif{t'}\e^{\i(E_{\kappa}-E_{c}+E_{b}+\i\eta)(t-t')/\hbar}\\
&\phantom{....}
\times\left(T_{cb_1}(\kappa)\Phi_{b_1b}^{[0]}(t')f_{\kappa}-\Phi_{cc_1}^{[0]}(t')T_{c_1b}(\kappa)f_{-\kappa}\right).
\end{aligned}
\end{equation}
After performing a \textit{Markov} approximation in the above integral, $\Phi_{bb'}^{[0]}(t')\approx\Phi_{bb'}^{[0]}(t)$, and setting $t\rightarrow+\infty$ we also obtain Eq.~\eqref{ssphi1_1vN}.
We note that there is another possibility to choose the time dependence of $\Phi^{[0]}$:
\begin{equation}\label{mark_R}
\Phi_{bb'}^{[0]}(t')\approx \e^{\i(E_{b}-E_{b'})(t-t')/\hbar}\Phi_{bb'}^{[0]}(t).
\end{equation}
In this case we obtain,
\begin{equation}\label{ssphi1_R}
\begin{aligned}
\Phi_{cb}^{[1]}(\kappa)&=
\frac{T_{cb_1}(\kappa)\Phi_{b_1b}^{[0]}f_{\kappa}}{E_{\kappa}-E_{c}+E_{b_1}+\i\eta}\\
&-\frac{\Phi_{cc_1}^{[0]}T_{c_1b}(\kappa)f_{-\kappa}}{E_{\kappa}-E_{c_1}+E_{b}+\i\eta}.
\end{aligned}
\end{equation}
which together with Eq.~\eqref{pheq0} provides the \textit{Wangsness-Bloch-Redfield} approach.~\cite{BreuerBook2006} The denominators in Eq.~\eqref{ssphi1_R} differ from Eq.~\eqref{ssphi1_1vN} if non-diagonal density matrix elements $\Phi_{bb'}^{[0]}$ are relevant. The oscillatory behavior in Eq.~\eqref{mark_R} is suggested by the first right-hand side term in Eq.~\eqref{pheq0}.

After combining Eqs.~\eqref{pheq0}, \eqref{statcond}, and \eqref{ssphi1_1vN} we get the 1vN approach equations for the steady state:
\begin{align}\label{ss_1vN}
&\begin{aligned}
0=&\Phi_{bb'}(E_{b}-E_{b'})\\
+&\sum_{b''\ell}\Phi_{bb''}\Big[\sum_{a}\Gamma_{b''a,ab'}^{\ell}I_{ba}^{\ell-}-\sum_{c}\Gamma_{b''c,cb'}^{\ell}I_{cb}^{\ell+*}\Big]\\
+&\sum_{b''\ell}\Phi_{b''b'}\Big[\sum_{c}\Gamma_{bc,cb''}^{\ell}I_{cb'}^{\ell+}-\sum_{a}\Gamma_{ba,ab''}^{\ell}I_{b'a}^{\ell-*}\Big]\\
+&\sum_{aa'\ell}\Phi_{aa'}\Gamma_{ba,a'b'}^{\ell}[I_{b'a}^{\ell+*}-I_{ba'}^{\ell+}]\\
+&\sum_{cc'\ell}\Phi_{cc'}\Gamma_{bc,c'b'}^{\ell}[I_{c'b}^{\ell-*}-I_{cb'}^{\ell-}].
\end{aligned}
\end{align}
Additionally, we impose the normalisation condition for diagonal density matrix elements:
\begin{equation}
\sum_{b}\Phi_{bb}=1.
\end{equation}
Here in Eq.~\eqref{ss_1vN} the tunneling rate matrix $\Gamma$ is defined as
\begin{equation}
\Gamma_{ba,a'b'}^{\ell}=2\pi\nu_{F}\sum_{\sigma}T_{ba}(\ell\sigma)T_{a'b'}(\ell\sigma),
\end{equation}
and the following integral was introduced
\begin{align}\label{ooxi}
2\pi I_{ba}^{\ell\pm}&=\mc{P}\int_{-D}^{D}\frac{\dif{E}f(\pm E)}{E-p_{ba}^{\ell}}
-\i\pi f(\pm p_{ba}^{\ell})\theta(D-\abs{p_{ba}^{\ell}}),\nonumber\\
p_{ba}^{\ell}&=E_{b}-E_{a}-\mu_{\ell},\\
f(E)&=(\exp[E/k_{\mr{B}}T]+1)^{-1},\nonumber
\end{align}
which appears after performing the $k$-sums using a flat density of states approximation, i.~e., $\sum_{k}\rightarrow \nu_{F}\int_{-D}^{D} \dif{E}$, with $\nu_{F}$ denoting the density of states at the Fermi level and $2D$ denoting the bandwidth of the leads. In our calculations we assume that the bandwidth of the leads is the largest energy scale. Note that, in the limit $D\rightarrow+\infty$ the results become bandwidth independent\cite{HaldanePRL1978} and that is what we also have checked in our numerical simulations. We note that the 1vN approach does not include any broadening effects of the quantum dot levels due to leads, which in principle could be included by different means (e.g., see Refs. [\onlinecite{PedersenPRB2005a,JinJCP2008,DordaPRB2014,ChenJPhysChemC2014}]).

Finally, we are interested in the current going from the lead $\ell$ into the quantum dots, which is given by
\begin{equation}\label{cureq}
\begin{aligned}
I_{\ell}(t)&=e\sum_{k\sigma}\frac{\pd}{\pd{t}}\avgs{\cd_{k\ell\sigma}(t)\can_{k\ell\sigma}(t)}\\
&=\frac{2e}{\hbar}\sum_{k\sigma}\Imag[T_{bc}(\ell\sigma)\Phi_{cb}^{[1]}(k\ell\sigma)].
\end{aligned}
\end{equation}
Here $A(t)=\e^{\i H t}A\e^{-\i H t}$ denotes the Heisenberg evolution of an operator $A$. In the steady state the current is obtained from Eq.~\eqref{cureq} in terms of $\Phi_{b'b}^{[0]}$ as
\begin{equation}
I_{\ell}=\frac{2e}{\hbar}\sum_{cb}\Imag\Big[
\sum_{b'}\Gamma_{bc,cb'}^{\ell}I_{cb}^{\ell+}\Phi_{b'b}^{[0]}
-\sum_{c'}\Gamma_{bc,c'b}^{\ell}I_{cb}^{\ell-}\Phi_{cc'}^{[0]}\Big],
\end{equation}
which is the main output presented in the paper. For all the calculations (except for \figurename~6) the principal part integral $\mc{P}$ in Eq.~\eqref{ooxi} is neglected. The motivation for neglecting these terms is given in \sectionname~\ref{sec:nprinp}.

\section{\label{sec:Pauli}Pauli master equation}
The Pauli master equation can be obtained from the 1vN approach by neglecting the coherences $\Phi_{bb'}$, $b\neq b'$. In such a case for the populations $P_{b}=\Phi_{bb}$ we obtain the equations:
\begin{align}
\begin{aligned}
&\sum_{a\ell}\left[P_{a}\Gamma_{a\rightarrow b}^{\ell}f(+p_{ba}^{\ell})-P_{b}\Gamma_{b\rightarrow a}^{\ell}f(-p_{ba}^{\ell})\right]\\
+&\sum_{c\ell}\left[P_{c}\Gamma_{c\rightarrow b}^{\ell}f(-p_{cb}^{\ell})-P_{b}\Gamma_{b\rightarrow c}^{\ell}f(+p_{cb}^{\ell})\right]=0,
\end{aligned}
\end{align}
where we have denoted $\Gamma_{a\rightarrow b}^{\ell}=\Gamma_{ab,ba}^{\ell}=\Gamma_{b\rightarrow a}^{\ell}=\Gamma_{ba,ab}^{\ell}$. Using the populations $P_{b}$ the steady states current is expressed as
\begin{equation}\label{current12}
I_{\ell}=\frac{e}{\hbar}\sum_{ab}[P_{a}\Gamma_{a\rightarrow b}^{\ell}f(+p_{ba}^{\ell})
-P_{b}\Gamma_{b\rightarrow a}^{\ell}f(-p_{ba}^{\ell})].
\end{equation}

\section{\label{sec:Parameters} Motivation for parameter values}

\begin{figure}
\includegraphics[width=0.95\columnwidth]{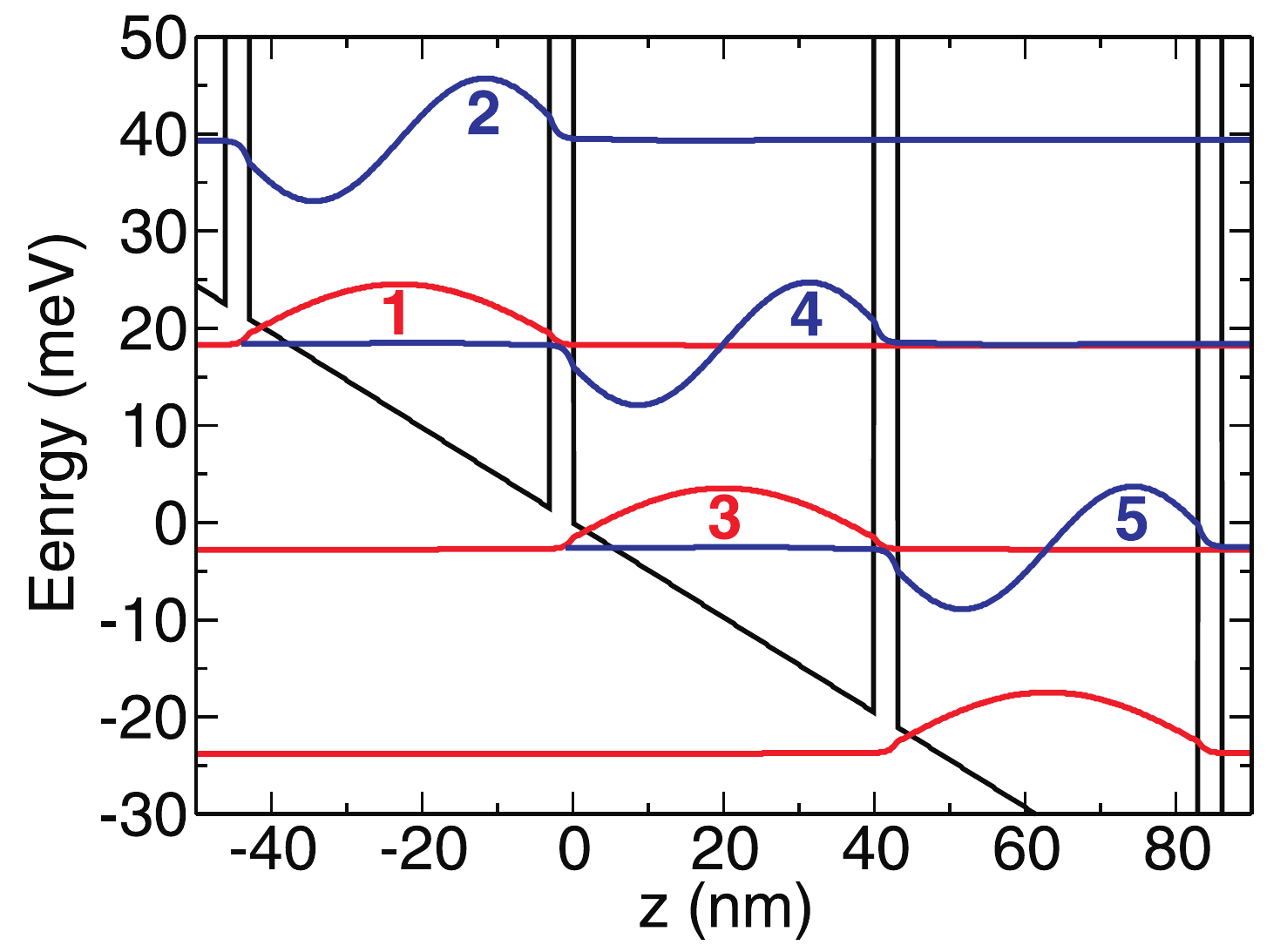}
\caption{\label{figS1} Wannier levels for the superlattice structure at a voltage drop of 21 meV per period. We use just three periods of this structure to model single particle levels in the serial triple dot. }
\end{figure}

We consider a superlattice structure with $40 \ \mathrm{nm}$ InAs wells ($m_\mathrm{eff}=0.026$) and $3 \ \mathrm{nm}$ InP barriers ($m_\mathrm{eff}=0.08$, $\Delta E_c=0.6 \ \mathrm{eV}$) providing the period $d=43 \ \mathrm{nm}$. The serial triple dot is described by just three periods of such a structure. A standard calculation (see, for example, Ref.~[\onlinecite{WackerPhysRep2002}]) provides the minibands $E_\nu(q)\approx E_\nu+2T_\nu\cos(qd) $ with
$E_a= 7 \ \mathrm{meV}$, $T_a=-0.058 \ \mathrm{meV}$,
$E_b= 28 \ \mathrm{meV}$, and $T_b=-0.23 \ \mathrm{meV}$.
Figure~\ref{figS1} displays the corresponding Wannier functions, where an electric field  of $21\mathrm{meV}/d$ is applied, so that the ground state $a$ of one well is in resonance with the excited state $b$ of the neighboring well.
The tunnel couplings between  Wannier states from the same band $\nu$
in neighboring periods is $T_\nu$. Thus we identify
\begin{itemize}
\item $\Omega_{13}=T_a=-0.058$ meV
\item $\Omega_{24}=\Omega_{45}=T_b=0.23$ meV
\end{itemize}
Furthermore, we obtain numerically the dipole matrix elements $s_{nm}=s_{mn}=\int\dif{z} \vphi_n^{*}(z)z \vphi_m(z)$ between the states. The non-vanishing values are $s_{21}=s_{43}=8.16$ nm (which is also used for the calculation of Coulomb matrix elements for the interdot interaction after making a dipole expansion, see Ref.~[\onlinecite{LassenPRB2007}]) and $s_{41}=s_{32}=s_{53}=-0.21$ nm. For the operation bias this  provides the tunnel couplings between states from different bands $\Omega_{nm}=-eFs_{nm}$ with the values
\begin{itemize}
\item $\Omega_{41}=\Omega_{53}=\Omega_{32}=0.103$ meV
\end{itemize}
We note, that the same procedure is used for simulating quantum cascade lasers, where we obtain quantitative agreement for a variety of samples.~\cite{LindskogAPL2014,DupontJAP2012}


\begin{thebibliography}{77}%
\makeatletter
\providecommand \@ifxundefined [1]{%
 \@ifx{#1\undefined}
}%
\providecommand \@ifnum [1]{%
 \ifnum #1\expandafter \@firstoftwo
 \else \expandafter \@secondoftwo
 \fi
}%
\providecommand \@ifx [1]{%
 \ifx #1\expandafter \@firstoftwo
 \else \expandafter \@secondoftwo
 \fi
}%
\providecommand \natexlab [1]{#1}%
\providecommand \enquote  [1]{``#1''}%
\providecommand \bibnamefont  [1]{#1}%
\providecommand \bibfnamefont [1]{#1}%
\providecommand \citenamefont [1]{#1}%
\providecommand \href@noop [0]{\@secondoftwo}%
\providecommand \href [0]{\begingroup \@sanitize@url \@href}%
\providecommand \@href[1]{\@@startlink{#1}\@@href}%
\providecommand \@@href[1]{\endgroup#1\@@endlink}%
\providecommand \@sanitize@url [0]{\catcode `\\12\catcode `\$12\catcode
  `\&12\catcode `\#12\catcode `\^12\catcode `\_12\catcode `\%12\relax}%
\providecommand \@@startlink[1]{}%
\providecommand \@@endlink[0]{}%
\providecommand \url  [0]{\begingroup\@sanitize@url \@url }%
\providecommand \@url [1]{\endgroup\@href {#1}{\urlprefix }}%
\providecommand \urlprefix  [0]{URL }%
\providecommand \Eprint [0]{\href }%
\providecommand \doibase [0]{http://dx.doi.org/}%
\providecommand \selectlanguage [0]{\@gobble}%
\providecommand \bibinfo  [0]{\@secondoftwo}%
\providecommand \bibfield  [0]{\@secondoftwo}%
\providecommand \translation [1]{[#1]}%
\providecommand \BibitemOpen [0]{}%
\providecommand \bibitemStop [0]{}%
\providecommand \bibitemNoStop [0]{.\EOS\space}%
\providecommand \EOS [0]{\spacefactor3000\relax}%
\providecommand \BibitemShut  [1]{\csname bibitem#1\endcsname}%
\let\auto@bib@innerbib\@empty
\bibitem [{\citenamefont {Devoret}\ \emph {et~al.}(1992)\citenamefont
  {Devoret}, \citenamefont {Esteve},\ and\ \citenamefont
  {Urbina}}]{DevoretNature1992}%
  \BibitemOpen
  \bibfield  {author} {\bibinfo {author} {\bibfnamefont {M.~H.}\ \bibnamefont
  {Devoret}}, \bibinfo {author} {\bibfnamefont {D.}~\bibnamefont {Esteve}}, \
  and\ \bibinfo {author} {\bibfnamefont {C.}~\bibnamefont {Urbina}},\ }\href
  {\doibase 10.1038/360547a0} {\bibfield  {journal} {\bibinfo  {journal}
  {Nature}\ }\textbf {\bibinfo {volume} {360}},\ \bibinfo {pages} {547}
  (\bibinfo {year} {1992})}\BibitemShut {NoStop}%
\bibitem [{\citenamefont {Tarucha}\ \emph {et~al.}(1996)\citenamefont
  {Tarucha}, \citenamefont {Austing}, \citenamefont {Honda}, \citenamefont
  {van~der Hage},\ and\ \citenamefont {Kouwenhoven}}]{TaruchaPRL1996}%
  \BibitemOpen
  \bibfield  {author} {\bibinfo {author} {\bibfnamefont {S.}~\bibnamefont
  {Tarucha}}, \bibinfo {author} {\bibfnamefont {D.~G.}\ \bibnamefont
  {Austing}}, \bibinfo {author} {\bibfnamefont {T.}~\bibnamefont {Honda}},
  \bibinfo {author} {\bibfnamefont {R.~J.}\ \bibnamefont {van~der Hage}}, \
  and\ \bibinfo {author} {\bibfnamefont {L.~P.}\ \bibnamefont {Kouwenhoven}},\
  }\href {\doibase 10.1103/PhysRevLett.77.3613} {\bibfield  {journal} {\bibinfo
   {journal} {Phys. Rev. Lett.}\ }\textbf {\bibinfo {volume} {77}},\ \bibinfo
  {pages} {3613} (\bibinfo {year} {1996})}\BibitemShut {NoStop}%
\bibitem [{\citenamefont {Kouwenhoven}\ \emph {et~al.}(1997)\citenamefont
  {Kouwenhoven}, \citenamefont {Oosterkamp}, \citenamefont {Danoesastro},
  \citenamefont {Eto}, \citenamefont {Austing}, \citenamefont {Honda},\ and\
  \citenamefont {Tarucha}}]{KouwenhovenScience1997}%
  \BibitemOpen
  \bibfield  {author} {\bibinfo {author} {\bibfnamefont {L.~P.}\ \bibnamefont
  {Kouwenhoven}}, \bibinfo {author} {\bibfnamefont {T.~H.}\ \bibnamefont
  {Oosterkamp}}, \bibinfo {author} {\bibfnamefont {M.~W.~S.}\ \bibnamefont
  {Danoesastro}}, \bibinfo {author} {\bibfnamefont {M.}~\bibnamefont {Eto}},
  \bibinfo {author} {\bibfnamefont {D.~G.}\ \bibnamefont {Austing}}, \bibinfo
  {author} {\bibfnamefont {T.}~\bibnamefont {Honda}}, \ and\ \bibinfo {author}
  {\bibfnamefont {S.}~\bibnamefont {Tarucha}},\ }\href {\doibase
  10.1126/science.278.5344.1788} {\bibfield  {journal} {\bibinfo  {journal}
  {Science}\ }\textbf {\bibinfo {volume} {278}},\ \bibinfo {pages} {1788}
  (\bibinfo {year} {1997})}\BibitemShut {NoStop}%
\bibitem [{\citenamefont {Reimann}\ and\ \citenamefont
  {Manninen}(2002)}]{ReimannRMP2002}%
  \BibitemOpen
  \bibfield  {author} {\bibinfo {author} {\bibfnamefont {S.~M.}\ \bibnamefont
  {Reimann}}\ and\ \bibinfo {author} {\bibfnamefont {M.}~\bibnamefont
  {Manninen}},\ }\href {\doibase http://dx.doi.org/10.1103/RevModPhys.74.1283}
  {\bibfield  {journal} {\bibinfo  {journal} {Rev.~Mod.~Phys.}\ }\textbf
  {\bibinfo {volume} {74}},\ \bibinfo {pages} {1283} (\bibinfo {year}
  {2002})}\BibitemShut {NoStop}%
\bibitem [{\citenamefont {Korkusinski}\ \emph {et~al.}(2007)\citenamefont
  {Korkusinski}, \citenamefont {Gimenez}, \citenamefont {Hawrylak},
  \citenamefont {Gaudreau}, \citenamefont {Studenikin},\ and\ \citenamefont
  {Sachrajda}}]{KorkusinskiPRB2007}%
  \BibitemOpen
  \bibfield  {author} {\bibinfo {author} {\bibfnamefont {M.}~\bibnamefont
  {Korkusinski}}, \bibinfo {author} {\bibfnamefont {I.~P.}\ \bibnamefont
  {Gimenez}}, \bibinfo {author} {\bibfnamefont {P.}~\bibnamefont {Hawrylak}},
  \bibinfo {author} {\bibfnamefont {L.}~\bibnamefont {Gaudreau}}, \bibinfo
  {author} {\bibfnamefont {S.~A.}\ \bibnamefont {Studenikin}}, \ and\ \bibinfo
  {author} {\bibfnamefont {A.~S.}\ \bibnamefont {Sachrajda}},\ }\href {\doibase
  10.1103/PhysRevB.75.115301} {\bibfield  {journal} {\bibinfo  {journal} {Phys.
  Rev. B}\ }\textbf {\bibinfo {volume} {75}},\ \bibinfo {pages} {115301}
  (\bibinfo {year} {2007})}\BibitemShut {NoStop}%
\bibitem [{\citenamefont {Kastner}(1992)}]{KastnerRMP1992}%
  \BibitemOpen
  \bibfield  {author} {\bibinfo {author} {\bibfnamefont {M.~A.}\ \bibnamefont
  {Kastner}},\ }\href {\doibase 10.1103/RevModPhys.64.849} {\bibfield
  {journal} {\bibinfo  {journal} {Rev. Mod. Phys.}\ }\textbf {\bibinfo {volume}
  {64}},\ \bibinfo {pages} {849} (\bibinfo {year} {1992})}\BibitemShut
  {NoStop}%
\bibitem [{\citenamefont {Averin}\ \emph {et~al.}(1991)\citenamefont {Averin},
  \citenamefont {Korotkov},\ and\ \citenamefont {Likharev}}]{AverinPRB1991}%
  \BibitemOpen
  \bibfield  {author} {\bibinfo {author} {\bibfnamefont {D.~V.}\ \bibnamefont
  {Averin}}, \bibinfo {author} {\bibfnamefont {A.~N.}\ \bibnamefont
  {Korotkov}}, \ and\ \bibinfo {author} {\bibfnamefont {K.~K.}\ \bibnamefont
  {Likharev}},\ }\href {\doibase 10.1103/PhysRevB.44.6199} {\bibfield
  {journal} {\bibinfo  {journal} {Phys.~Rev.~B}\ }\textbf {\bibinfo {volume}
  {44}},\ \bibinfo {pages} {6199} (\bibinfo {year} {1991})}\BibitemShut
  {NoStop}%
\bibitem [{\citenamefont {Fulton}\ and\ \citenamefont
  {Dolan}(1987)}]{FultonPRL1987}%
  \BibitemOpen
  \bibfield  {author} {\bibinfo {author} {\bibfnamefont {T.~A.}\ \bibnamefont
  {Fulton}}\ and\ \bibinfo {author} {\bibfnamefont {G.~J.}\ \bibnamefont
  {Dolan}},\ }\href {\doibase 10.1103/PhysRevLett.59.109} {\bibfield  {journal}
  {\bibinfo  {journal} {Phys. Rev. Lett.}\ }\textbf {\bibinfo {volume} {59}},\
  \bibinfo {pages} {109} (\bibinfo {year} {1987})}\BibitemShut {NoStop}%
\bibitem [{\citenamefont {Ashoori}(1996)}]{AshooriNature1996}%
  \BibitemOpen
  \bibfield  {author} {\bibinfo {author} {\bibfnamefont {R.~C.}\ \bibnamefont
  {Ashoori}},\ }\href {\doibase 10.1038/379413a0} {\bibfield  {journal}
  {\bibinfo  {journal} {Nature}\ }\textbf {\bibinfo {volume} {379}},\ \bibinfo
  {pages} {413} (\bibinfo {year} {1996})}\BibitemShut {NoStop}%
\bibitem [{\citenamefont {Kouwenhoven}\ \emph {et~al.}(1991)\citenamefont
  {Kouwenhoven}, \citenamefont {{van der}~Vaart}, \citenamefont {Johnson},
  \citenamefont {Kool}, \citenamefont {Harmans}, \citenamefont {Williamson},
  \citenamefont {Staring},\ and\ \citenamefont
  {Foxon}}]{KouwenhovenZPhysBCondMat1991}%
  \BibitemOpen
  \bibfield  {author} {\bibinfo {author} {\bibfnamefont {L.~P.}\ \bibnamefont
  {Kouwenhoven}}, \bibinfo {author} {\bibfnamefont {N.~C.}\ \bibnamefont {{van
  der}~Vaart}}, \bibinfo {author} {\bibfnamefont {A.~T.}\ \bibnamefont
  {Johnson}}, \bibinfo {author} {\bibfnamefont {W.}~\bibnamefont {Kool}},
  \bibinfo {author} {\bibfnamefont {C.~J. P.~M.}\ \bibnamefont {Harmans}},
  \bibinfo {author} {\bibfnamefont {J.~G.}\ \bibnamefont {Williamson}},
  \bibinfo {author} {\bibfnamefont {A.~A.~M.}\ \bibnamefont {Staring}}, \ and\
  \bibinfo {author} {\bibfnamefont {C.~T.}\ \bibnamefont {Foxon}},\ }\href
  {\doibase 10.1007/BF01307632} {\bibfield  {journal} {\bibinfo  {journal} {Z.
  Phys. B - Con. Mat.}\ }\textbf {\bibinfo {volume} {85}},\ \bibinfo {pages}
  {367} (\bibinfo {year} {1991})}\BibitemShut {NoStop}%
\bibitem [{\citenamefont {Averin}\ and\ \citenamefont
  {Likharev}(1986)}]{AverinJLTP1986}%
  \BibitemOpen
  \bibfield  {author} {\bibinfo {author} {\bibfnamefont {D.~V.}\ \bibnamefont
  {Averin}}\ and\ \bibinfo {author} {\bibfnamefont {K.~K.}\ \bibnamefont
  {Likharev}},\ }\href {\doibase 10.1007/BF00683469} {\bibfield  {journal}
  {\bibinfo  {journal} {J. Low. Temp. Phys.}\ }\textbf {\bibinfo {volume}
  {62}},\ \bibinfo {pages} {345} (\bibinfo {year} {1986})}\BibitemShut
  {NoStop}%
\bibitem [{\citenamefont {Yoffe}(2001)}]{YoffeAdvancesinPhysics2001}%
  \BibitemOpen
  \bibfield  {author} {\bibinfo {author} {\bibfnamefont {A.~D.}\ \bibnamefont
  {Yoffe}},\ }\href {\doibase 10.1080/00018730010006608} {\bibfield  {journal}
  {\bibinfo  {journal} {Adv. Phys.}\ }\textbf {\bibinfo {volume} {50}},\
  \bibinfo {pages} {1} (\bibinfo {year} {2001})}\BibitemShut {NoStop}%
\bibitem [{\citenamefont {Beard}(2011)}]{BeardPCL2011}%
  \BibitemOpen
  \bibfield  {author} {\bibinfo {author} {\bibfnamefont {M.~C.}\ \bibnamefont
  {Beard}},\ }\href {\doibase 10.1021/jz200166y} {\bibfield  {journal}
  {\bibinfo  {journal} {J. Phys. Chem. Lett.}\ }\textbf {\bibinfo {volume}
  {2}},\ \bibinfo {pages} {1282} (\bibinfo {year} {2011})}\BibitemShut
  {NoStop}%
\bibitem [{\citenamefont {Klimov}(2007)}]{KlimovARPC2007}%
  \BibitemOpen
  \bibfield  {author} {\bibinfo {author} {\bibfnamefont {V.~I.}\ \bibnamefont
  {Klimov}},\ }\href {\doibase 10.1146/annurev.physchem.58.032806.104537}
  {\bibfield  {journal} {\bibinfo  {journal} {Annu. Rev. Phys. Chem.}\ }\textbf
  {\bibinfo {volume} {58}},\ \bibinfo {pages} {635} (\bibinfo {year}
  {2007})}\BibitemShut {NoStop}%
\bibitem [{\citenamefont {Waugh}\ \emph {et~al.}(1995)\citenamefont {Waugh},
  \citenamefont {Berry}, \citenamefont {Mar}, \citenamefont {Westervelt},
  \citenamefont {Campman},\ and\ \citenamefont {Gossard}}]{WaughPRL1995}%
  \BibitemOpen
  \bibfield  {author} {\bibinfo {author} {\bibfnamefont {F.}~\bibnamefont
  {Waugh}}, \bibinfo {author} {\bibfnamefont {M.}~\bibnamefont {Berry}},
  \bibinfo {author} {\bibfnamefont {D.}~\bibnamefont {Mar}}, \bibinfo {author}
  {\bibfnamefont {R.}~\bibnamefont {Westervelt}}, \bibinfo {author}
  {\bibfnamefont {K.}~\bibnamefont {Campman}}, \ and\ \bibinfo {author}
  {\bibfnamefont {A.}~\bibnamefont {Gossard}},\ }\href {\doibase
  10.1103/PhysRevLett.75.705} {\bibfield  {journal} {\bibinfo  {journal} {Phys.
  Rev. Lett.}\ }\textbf {\bibinfo {volume} {75}},\ \bibinfo {pages} {705}
  (\bibinfo {year} {1995})}\BibitemShut {NoStop}%
\bibitem [{\citenamefont {Fujisawa}\ \emph {et~al.}(1998)\citenamefont
  {Fujisawa}, \citenamefont {Oosterkamp}, \citenamefont {van~der Wiel},
  \citenamefont {Broer}, \citenamefont {Aguado}, \citenamefont {Tarucha},\ and\
  \citenamefont {Kouwenho~ven}}]{FujisawaScience1998}%
  \BibitemOpen
  \bibfield  {author} {\bibinfo {author} {\bibfnamefont {T.}~\bibnamefont
  {Fujisawa}}, \bibinfo {author} {\bibfnamefont {T.~H.}\ \bibnamefont
  {Oosterkamp}}, \bibinfo {author} {\bibfnamefont {W.~G.}\ \bibnamefont
  {van~der Wiel}}, \bibinfo {author} {\bibfnamefont {B.~W.}\ \bibnamefont
  {Broer}}, \bibinfo {author} {\bibfnamefont {R.}~\bibnamefont {Aguado}},
  \bibinfo {author} {\bibfnamefont {S.}~\bibnamefont {Tarucha}}, \ and\
  \bibinfo {author} {\bibfnamefont {L.~P.}\ \bibnamefont {Kouwenho~ven}},\
  }\href {\doibase 10.1126/science.282.5390.932} {\bibfield  {journal}
  {\bibinfo  {journal} {Science}\ }\textbf {\bibinfo {volume} {282}},\ \bibinfo
  {pages} {932} (\bibinfo {year} {1998})}\BibitemShut {NoStop}%
\bibitem [{\citenamefont {Liu}\ \emph {et~al.}(2010)\citenamefont {Liu},
  \citenamefont {Hoffman}, \citenamefont {Escarra}, \citenamefont {Franz},
  \citenamefont {Khurgin}, \citenamefont {Dikmelik}, \citenamefont {Wang},
  \citenamefont {Fan},\ and\ \citenamefont {Gmachl}}]{LiuNatPhot2010}%
  \BibitemOpen
  \bibfield  {author} {\bibinfo {author} {\bibfnamefont {P.~Q.}\ \bibnamefont
  {Liu}}, \bibinfo {author} {\bibfnamefont {A.~J.}\ \bibnamefont {Hoffman}},
  \bibinfo {author} {\bibfnamefont {M.~D.}\ \bibnamefont {Escarra}}, \bibinfo
  {author} {\bibfnamefont {K.~J.}\ \bibnamefont {Franz}}, \bibinfo {author}
  {\bibfnamefont {J.~B.}\ \bibnamefont {Khurgin}}, \bibinfo {author}
  {\bibfnamefont {Y.}~\bibnamefont {Dikmelik}}, \bibinfo {author}
  {\bibfnamefont {X.}~\bibnamefont {Wang}}, \bibinfo {author} {\bibfnamefont
  {J.}~\bibnamefont {Fan}}, \ and\ \bibinfo {author} {\bibfnamefont {C.~F.}\
  \bibnamefont {Gmachl}},\ }\href@noop {} {\bibfield  {journal} {\bibinfo
  {journal} {Nat.~Photonics}\ }\textbf {\bibinfo {volume} {4}},\ \bibinfo
  {pages} {95} (\bibinfo {year} {2010})}\BibitemShut {NoStop}%
\bibitem [{\citenamefont {Schedelbeck}\ \emph {et~al.}(1997)\citenamefont
  {Schedelbeck}, \citenamefont {Wegscheider}, \citenamefont {Bichler},\ and\
  \citenamefont {Abstreiter}}]{SchedelbeckScience1997}%
  \BibitemOpen
  \bibfield  {author} {\bibinfo {author} {\bibfnamefont {G.}~\bibnamefont
  {Schedelbeck}}, \bibinfo {author} {\bibfnamefont {W.}~\bibnamefont
  {Wegscheider}}, \bibinfo {author} {\bibfnamefont {M.}~\bibnamefont
  {Bichler}}, \ and\ \bibinfo {author} {\bibfnamefont {G.}~\bibnamefont
  {Abstreiter}},\ }\href@noop {} {\bibfield  {journal} {\bibinfo  {journal}
  {Science}\ }\textbf {\bibinfo {volume} {278}},\ \bibinfo {pages} {1792}
  (\bibinfo {year} {1997})}\BibitemShut {NoStop}%
\bibitem [{\citenamefont {Borgstrom}\ \emph {et~al.}(2001)\citenamefont
  {Borgstrom}, \citenamefont {Bryllert}, \citenamefont {Sass}, \citenamefont
  {Gustafson}, \citenamefont {Wernersson}, \citenamefont {Seifert},\ and\
  \citenamefont {Samuelson}}]{BorgstromAPL2001}%
  \BibitemOpen
  \bibfield  {author} {\bibinfo {author} {\bibfnamefont {M.}~\bibnamefont
  {Borgstrom}}, \bibinfo {author} {\bibfnamefont {T.}~\bibnamefont {Bryllert}},
  \bibinfo {author} {\bibfnamefont {T.}~\bibnamefont {Sass}}, \bibinfo {author}
  {\bibfnamefont {B.}~\bibnamefont {Gustafson}}, \bibinfo {author}
  {\bibfnamefont {L.-E.}\ \bibnamefont {Wernersson}}, \bibinfo {author}
  {\bibfnamefont {W.}~\bibnamefont {Seifert}}, \ and\ \bibinfo {author}
  {\bibfnamefont {L.}~\bibnamefont {Samuelson}},\ }\href {\doibase
  http://dx.doi.org/10.1063/1.1374235} {\bibfield  {journal} {\bibinfo
  {journal} {Appl.~Phys.~Lett.}\ }\textbf {\bibinfo {volume} {78}},\ \bibinfo
  {pages} {3232} (\bibinfo {year} {2001})}\BibitemShut {NoStop}%
\bibitem [{\citenamefont {Mason}\ \emph {et~al.}(2004)\citenamefont {Mason},
  \citenamefont {Biercuk},\ and\ \citenamefont {Marcus}}]{MasonScience2004}%
  \BibitemOpen
  \bibfield  {author} {\bibinfo {author} {\bibfnamefont {N.}~\bibnamefont
  {Mason}}, \bibinfo {author} {\bibfnamefont {M.~J.}\ \bibnamefont {Biercuk}},
  \ and\ \bibinfo {author} {\bibfnamefont {C.~M.}\ \bibnamefont {Marcus}},\
  }\href {\doibase 10.1126/science.1093605} {\bibfield  {journal} {\bibinfo
  {journal} {Science}\ }\textbf {\bibinfo {volume} {303}},\ \bibinfo {pages}
  {655} (\bibinfo {year} {2004})}\BibitemShut {NoStop}%
\bibitem [{\citenamefont {Fasth}\ \emph {et~al.}(2005)\citenamefont {Fasth},
  \citenamefont {Fuhrer}, \citenamefont {Bj{\"o}rk},\ and\ \citenamefont
  {Samuelson}}]{FasthNL2005}%
  \BibitemOpen
  \bibfield  {author} {\bibinfo {author} {\bibfnamefont {C.}~\bibnamefont
  {Fasth}}, \bibinfo {author} {\bibfnamefont {A.}~\bibnamefont {Fuhrer}},
  \bibinfo {author} {\bibfnamefont {M.~T.}\ \bibnamefont {Bj{\"o}rk}}, \ and\
  \bibinfo {author} {\bibfnamefont {L.}~\bibnamefont {Samuelson}},\ }\href
  {\doibase 10.1021/nl050850i} {\bibfield  {journal} {\bibinfo  {journal} {Nano
  Lett.}\ }\textbf {\bibinfo {volume} {5}},\ \bibinfo {pages} {1487} (\bibinfo
  {year} {2005})}\BibitemShut {NoStop}%
\bibitem [{\citenamefont {Fuhrer}\ \emph {et~al.}(2007)\citenamefont {Fuhrer},
  \citenamefont {Fr{\"o}berg}, \citenamefont {Pedersen}, \citenamefont
  {Larsson}, \citenamefont {Wacker}, \citenamefont {Pistol},\ and\
  \citenamefont {Samuelson}}]{FuhrerNL2007}%
  \BibitemOpen
  \bibfield  {author} {\bibinfo {author} {\bibfnamefont {A.}~\bibnamefont
  {Fuhrer}}, \bibinfo {author} {\bibfnamefont {L.~E.}\ \bibnamefont
  {Fr{\"o}berg}}, \bibinfo {author} {\bibfnamefont {J.~N.}\ \bibnamefont
  {Pedersen}}, \bibinfo {author} {\bibfnamefont {M.~W.}\ \bibnamefont
  {Larsson}}, \bibinfo {author} {\bibfnamefont {A.}~\bibnamefont {Wacker}},
  \bibinfo {author} {\bibfnamefont {M.-E.}\ \bibnamefont {Pistol}}, \ and\
  \bibinfo {author} {\bibfnamefont {L.}~\bibnamefont {Samuelson}},\ }\href
  {\doibase 10.1021/nl061913f} {\bibfield  {journal} {\bibinfo  {journal} {Nano
  Lett.}\ }\textbf {\bibinfo {volume} {7}},\ \bibinfo {pages} {243} (\bibinfo
  {year} {2007})}\BibitemShut {NoStop}%
\bibitem [{\citenamefont {F{\"o}lsch}\ \emph {et~al.}(2014)\citenamefont
  {F{\"o}lsch}, \citenamefont {Mart{\'\i}nez-Blanco}, \citenamefont {Yang},
  \citenamefont {Kanisawa},\ and\ \citenamefont
  {Erwin}}]{FolschNatNanotechnol2014}%
  \BibitemOpen
  \bibfield  {author} {\bibinfo {author} {\bibfnamefont {S.}~\bibnamefont
  {F{\"o}lsch}}, \bibinfo {author} {\bibfnamefont {J.}~\bibnamefont
  {Mart{\'\i}nez-Blanco}}, \bibinfo {author} {\bibfnamefont {J.}~\bibnamefont
  {Yang}}, \bibinfo {author} {\bibfnamefont {K.}~\bibnamefont {Kanisawa}}, \
  and\ \bibinfo {author} {\bibfnamefont {S.~C.}\ \bibnamefont {Erwin}},\
  }\href@noop {} {\bibfield  {journal} {\bibinfo  {journal} {Nat.
  Nanotechnol.}\ }\textbf {\bibinfo {volume} {9}},\ \bibinfo {pages} {505}
  (\bibinfo {year} {2014})}\BibitemShut {NoStop}%
\bibitem [{\citenamefont {Loss}\ and\ \citenamefont
  {{DiVincenzo}}(1998)}]{LossPRA1998}%
  \BibitemOpen
  \bibfield  {author} {\bibinfo {author} {\bibfnamefont {D.}~\bibnamefont
  {Loss}}\ and\ \bibinfo {author} {\bibfnamefont {D.~P.}\ \bibnamefont
  {{DiVincenzo}}},\ }\href {\doibase 10.1103/PhysRevA.57.120} {\bibfield
  {journal} {\bibinfo  {journal} {Phys. Rev. A}\ }\textbf {\bibinfo {volume}
  {57}},\ \bibinfo {pages} {120} (\bibinfo {year} {1998})}\BibitemShut
  {NoStop}%
\bibitem [{\citenamefont {Laird}\ \emph {et~al.}(2010)\citenamefont {Laird},
  \citenamefont {Taylor}, \citenamefont {{DiVincenzo}}, \citenamefont {Marcus},
  \citenamefont {Hanson},\ and\ \citenamefont {Gossard}}]{LairdPRB2010}%
  \BibitemOpen
  \bibfield  {author} {\bibinfo {author} {\bibfnamefont {E.~A.}\ \bibnamefont
  {Laird}}, \bibinfo {author} {\bibfnamefont {J.~M.}\ \bibnamefont {Taylor}},
  \bibinfo {author} {\bibfnamefont {D.~P.}\ \bibnamefont {{DiVincenzo}}},
  \bibinfo {author} {\bibfnamefont {C.~M.}\ \bibnamefont {Marcus}}, \bibinfo
  {author} {\bibfnamefont {M.~P.}\ \bibnamefont {Hanson}}, \ and\ \bibinfo
  {author} {\bibfnamefont {A.~C.}\ \bibnamefont {Gossard}},\ }\href {\doibase
  http://dx.doi.org/10.1103/PhysRevB.82.075403} {\bibfield  {journal} {\bibinfo
   {journal} {Phys. Rev. B}\ }\textbf {\bibinfo {volume} {82}},\ \bibinfo
  {pages} {075403} (\bibinfo {year} {2010})}\BibitemShut {NoStop}%
\bibitem [{\citenamefont {Burnett}\ and\ \citenamefont
  {Williams}(2014)}]{BurnettPRB2014}%
  \BibitemOpen
  \bibfield  {author} {\bibinfo {author} {\bibfnamefont {B.~A.}\ \bibnamefont
  {Burnett}}\ and\ \bibinfo {author} {\bibfnamefont {B.~S.}\ \bibnamefont
  {Williams}},\ }\href {\doibase 10.1103/PhysRevB.90.155309} {\bibfield
  {journal} {\bibinfo  {journal} {Phys. Rev. B}\ }\textbf {\bibinfo {volume}
  {90}},\ \bibinfo {pages} {155309} (\bibinfo {year} {2014})}\BibitemShut
  {NoStop}%
\bibitem [{\citenamefont {Grange}(2014{\natexlab{a}})}]{GrangeAPL2014}%
  \BibitemOpen
  \bibfield  {author} {\bibinfo {author} {\bibfnamefont {T.}~\bibnamefont
  {Grange}},\ }\href {\doibase 10.1063/1.4897543} {\bibfield  {journal}
  {\bibinfo  {journal} {Appl. Phys. Lett.}\ }\textbf {\bibinfo {volume}
  {105}},\ \bibinfo {pages} {141105} (\bibinfo {year}
  {2014}{\natexlab{a}})}\BibitemShut {NoStop}%
\bibitem [{\citenamefont {van~der Wiel}\ \emph {et~al.}(2003)\citenamefont
  {van~der Wiel}, \citenamefont {De~Franceschi}, \citenamefont {Elzerman},
  \citenamefont {Fujisawa}, \citenamefont {Tarucha},\ and\ \citenamefont
  {Kouwenhoven}}]{VanDerWielRMP2003}%
  \BibitemOpen
  \bibfield  {author} {\bibinfo {author} {\bibfnamefont {W.~G.}\ \bibnamefont
  {van~der Wiel}}, \bibinfo {author} {\bibfnamefont {S.}~\bibnamefont
  {De~Franceschi}}, \bibinfo {author} {\bibfnamefont {J.~M.}\ \bibnamefont
  {Elzerman}}, \bibinfo {author} {\bibfnamefont {T.}~\bibnamefont {Fujisawa}},
  \bibinfo {author} {\bibfnamefont {S.}~\bibnamefont {Tarucha}}, \ and\
  \bibinfo {author} {\bibfnamefont {L.~P.}\ \bibnamefont {Kouwenhoven}},\
  }\href {\doibase 10.1103/RevModPhys.75.1} {\bibfield  {journal} {\bibinfo
  {journal} {Rev.~Mod.~Phys.}\ }\textbf {\bibinfo {volume} {75}},\ \bibinfo
  {pages} {1} (\bibinfo {year} {2003})}\BibitemShut {NoStop}%
\bibitem [{\citenamefont {Hanson}\ \emph {et~al.}(2007)\citenamefont {Hanson},
  \citenamefont {Kouwenhoven}, \citenamefont {Petta}, \citenamefont {Tarucha},\
  and\ \citenamefont {Vandersypen}}]{HansonRMP2007}%
  \BibitemOpen
  \bibfield  {author} {\bibinfo {author} {\bibfnamefont {R.}~\bibnamefont
  {Hanson}}, \bibinfo {author} {\bibfnamefont {L.~P.}\ \bibnamefont
  {Kouwenhoven}}, \bibinfo {author} {\bibfnamefont {J.~R.}\ \bibnamefont
  {Petta}}, \bibinfo {author} {\bibfnamefont {S.}~\bibnamefont {Tarucha}}, \
  and\ \bibinfo {author} {\bibfnamefont {L.~M.~K.}\ \bibnamefont
  {Vandersypen}},\ }\href {\doibase 10.1103/RevModPhys.79.1217} {\bibfield
  {journal} {\bibinfo  {journal} {Rev. Mod. Phys.}\ }\textbf {\bibinfo {volume}
  {79}},\ \bibinfo {pages} {1217} (\bibinfo {year} {2007})}\BibitemShut
  {NoStop}%
\bibitem [{\citenamefont {Jespersen}\ \emph {et~al.}(2011)\citenamefont
  {Jespersen}, \citenamefont {Grove-Rasmussen}, \citenamefont {Paaske},
  \citenamefont {Muraki}, \citenamefont {Fujisawa}, \citenamefont {Nyg\aa~rd},\
  and\ \citenamefont {Flensberg}}]{JespersenNatPhys2011}%
  \BibitemOpen
  \bibfield  {author} {\bibinfo {author} {\bibfnamefont {T.~S.}\ \bibnamefont
  {Jespersen}}, \bibinfo {author} {\bibfnamefont {K.}~\bibnamefont
  {Grove-Rasmussen}}, \bibinfo {author} {\bibfnamefont {J.}~\bibnamefont
  {Paaske}}, \bibinfo {author} {\bibfnamefont {K.}~\bibnamefont {Muraki}},
  \bibinfo {author} {\bibfnamefont {T.}~\bibnamefont {Fujisawa}}, \bibinfo
  {author} {\bibfnamefont {J.}~\bibnamefont {Nyg\aa~rd}}, \ and\ \bibinfo
  {author} {\bibfnamefont {K.}~\bibnamefont {Flensberg}},\ }\href {\doibase
  10.1038/nphys1880} {\bibfield  {journal} {\bibinfo  {journal} {Nat. Phys.}\
  }\textbf {\bibinfo {volume} {7}},\ \bibinfo {pages} {348} (\bibinfo {year}
  {2011})}\BibitemShut {NoStop}%
\bibitem [{\citenamefont {Hatano}\ \emph {et~al.}(2013)\citenamefont {Hatano},
  \citenamefont {Tokura}, \citenamefont {Amaha}, \citenamefont {Kubo},
  \citenamefont {Teraoka},\ and\ \citenamefont {Tarucha}}]{HatanoPRB2013}%
  \BibitemOpen
  \bibfield  {author} {\bibinfo {author} {\bibfnamefont {T.}~\bibnamefont
  {Hatano}}, \bibinfo {author} {\bibfnamefont {Y.}~\bibnamefont {Tokura}},
  \bibinfo {author} {\bibfnamefont {S.}~\bibnamefont {Amaha}}, \bibinfo
  {author} {\bibfnamefont {T.}~\bibnamefont {Kubo}}, \bibinfo {author}
  {\bibfnamefont {S.}~\bibnamefont {Teraoka}}, \ and\ \bibinfo {author}
  {\bibfnamefont {S.}~\bibnamefont {Tarucha}},\ }\href {\doibase
  10.1103/PhysRevB.87.241414} {\bibfield  {journal} {\bibinfo  {journal} {Phys.
  Rev. B}\ }\textbf {\bibinfo {volume} {87}},\ \bibinfo {pages} {241414}
  (\bibinfo {year} {2013})}\BibitemShut {NoStop}%
\bibitem [{\citenamefont {Brandes}\ and\ \citenamefont
  {Kramer}(1999)}]{BrandesPRL1999}%
  \BibitemOpen
  \bibfield  {author} {\bibinfo {author} {\bibfnamefont {T.}~\bibnamefont
  {Brandes}}\ and\ \bibinfo {author} {\bibfnamefont {B.}~\bibnamefont
  {Kramer}},\ }\href@noop {} {\bibfield  {journal} {\bibinfo  {journal}
  {Phys.~Rev.~Lett.}\ }\textbf {\bibinfo {volume} {83}},\ \bibinfo {pages}
  {3021} (\bibinfo {year} {1999})}\BibitemShut {NoStop}%
\bibitem [{\citenamefont {Weber}\ \emph {et~al.}(2010)\citenamefont {Weber},
  \citenamefont {Fuhrer}, \citenamefont {Fasth}, \citenamefont {Lindwall},
  \citenamefont {Samuelson},\ and\ \citenamefont {Wacker}}]{WeberPRL2010}%
  \BibitemOpen
  \bibfield  {author} {\bibinfo {author} {\bibfnamefont {C.}~\bibnamefont
  {Weber}}, \bibinfo {author} {\bibfnamefont {A.}~\bibnamefont {Fuhrer}},
  \bibinfo {author} {\bibfnamefont {C.}~\bibnamefont {Fasth}}, \bibinfo
  {author} {\bibfnamefont {G.}~\bibnamefont {Lindwall}}, \bibinfo {author}
  {\bibfnamefont {L.}~\bibnamefont {Samuelson}}, \ and\ \bibinfo {author}
  {\bibfnamefont {A.}~\bibnamefont {Wacker}},\ }\href {\doibase
  10.1103/PhysRevLett.104.036801} {\bibfield  {journal} {\bibinfo  {journal}
  {Phys.~Rev.~Lett.}\ }\textbf {\bibinfo {volume} {104}},\ \bibinfo {pages}
  {036801} (\bibinfo {year} {2010})}\BibitemShut {NoStop}%
\bibitem [{\citenamefont {Pedersen}\ \emph {et~al.}(2007)\citenamefont
  {Pedersen}, \citenamefont {Lassen}, \citenamefont {Wacker},\ and\
  \citenamefont {Hettler}}]{PedersenPRB2007}%
  \BibitemOpen
  \bibfield  {author} {\bibinfo {author} {\bibfnamefont {J.~N.}\ \bibnamefont
  {Pedersen}}, \bibinfo {author} {\bibfnamefont {B.}~\bibnamefont {Lassen}},
  \bibinfo {author} {\bibfnamefont {A.}~\bibnamefont {Wacker}}, \ and\ \bibinfo
  {author} {\bibfnamefont {M.~H.}\ \bibnamefont {Hettler}},\ }\href {\doibase
  10.1103/PhysRevB.75.235314} {\bibfield  {journal} {\bibinfo  {journal}
  {Phys.~Rev.~B}\ }\textbf {\bibinfo {volume} {75}},\ \bibinfo {pages} {235314}
  (\bibinfo {year} {2007})}\BibitemShut {NoStop}%
\bibitem [{\citenamefont {Weymann}\ \emph {et~al.}(2011)\citenamefont
  {Weymann}, \citenamefont {Bu\l{}ka},\ and\ \citenamefont
  {Barna\ifmmode~\acute{s}\else \'{s}\fi{}}}]{WeymannPRB2011}%
  \BibitemOpen
  \bibfield  {author} {\bibinfo {author} {\bibfnamefont {I.}~\bibnamefont
  {Weymann}}, \bibinfo {author} {\bibfnamefont {B.~R.}\ \bibnamefont
  {Bu\l{}ka}}, \ and\ \bibinfo {author} {\bibfnamefont {J.}~\bibnamefont
  {Barna\ifmmode~\acute{s}\else \'{s}\fi{}}},\ }\href {\doibase
  10.1103/PhysRevB.83.195302} {\bibfield  {journal} {\bibinfo  {journal} {Phys.
  Rev. B}\ }\textbf {\bibinfo {volume} {83}},\ \bibinfo {pages} {195302}
  (\bibinfo {year} {2011})}\BibitemShut {NoStop}%
\bibitem [{\citenamefont {Emary}(2007)}]{EmaryPRB2007}%
  \BibitemOpen
  \bibfield  {author} {\bibinfo {author} {\bibfnamefont {C.}~\bibnamefont
  {Emary}},\ }\href {\doibase 10.1103/PhysRevB.76.245319} {\bibfield  {journal}
  {\bibinfo  {journal} {Phys. Rev. B}\ }\textbf {\bibinfo {volume} {76}},\
  \bibinfo {pages} {245319} (\bibinfo {year} {2007})}\BibitemShut {NoStop}%
\bibitem [{\citenamefont {Faist}(2013)}]{FaistBook2013}%
  \BibitemOpen
  \bibfield  {author} {\bibinfo {author} {\bibfnamefont {J.}~\bibnamefont
  {Faist}},\ }\href@noop {} {\emph {\bibinfo {title} {Quantum Cascade
  Lasers}}}\ (\bibinfo  {publisher} {Oxford University Press},\ \bibinfo
  {address} {Oxford},\ \bibinfo {year} {2013})\BibitemShut {NoStop}%
\bibitem [{\citenamefont {Grange}(2014{\natexlab{b}})}]{GrangePRB2014}%
  \BibitemOpen
  \bibfield  {author} {\bibinfo {author} {\bibfnamefont {T.}~\bibnamefont
  {Grange}},\ }\href {\doibase 10.1103/PhysRevB.89.165310} {\bibfield
  {journal} {\bibinfo  {journal} {Phys. Rev. B}\ }\textbf {\bibinfo {volume}
  {89}},\ \bibinfo {pages} {165310} (\bibinfo {year}
  {2014}{\natexlab{b}})}\BibitemShut {NoStop}%
\bibitem [{\citenamefont {Schr\"oer}\ \emph {et~al.}(2007)\citenamefont
  {Schr\"oer}, \citenamefont {Greentree}, \citenamefont {Gaudreau},
  \citenamefont {Eberl}, \citenamefont {Hollenberg}, \citenamefont {Kotthaus},\
  and\ \citenamefont {Ludwig}}]{SchroerPRB2007}%
  \BibitemOpen
  \bibfield  {author} {\bibinfo {author} {\bibfnamefont {D.}~\bibnamefont
  {Schr\"oer}}, \bibinfo {author} {\bibfnamefont {A.}~\bibnamefont
  {Greentree}}, \bibinfo {author} {\bibfnamefont {L.}~\bibnamefont {Gaudreau}},
  \bibinfo {author} {\bibfnamefont {K.}~\bibnamefont {Eberl}}, \bibinfo
  {author} {\bibfnamefont {L.}~\bibnamefont {Hollenberg}}, \bibinfo {author}
  {\bibfnamefont {J.}~\bibnamefont {Kotthaus}}, \ and\ \bibinfo {author}
  {\bibfnamefont {S.}~\bibnamefont {Ludwig}},\ }\href {\doibase
  10.1103/PhysRevB.76.075306} {\bibfield  {journal} {\bibinfo  {journal} {Phys.
  Rev. B}\ }\textbf {\bibinfo {volume} {76}},\ \bibinfo {pages} {075306}
  (\bibinfo {year} {2007})}\BibitemShut {NoStop}%
\bibitem [{\citenamefont {Grove-Rasmussen}\ \emph {et~al.}(2008)\citenamefont
  {Grove-Rasmussen}, \citenamefont {J{\o}rgensen}, \citenamefont {Hayashi},
  \citenamefont {Lindelof},\ and\ \citenamefont {Fujisawa}}]{GroveNL2008}%
  \BibitemOpen
  \bibfield  {author} {\bibinfo {author} {\bibfnamefont {K.}~\bibnamefont
  {Grove-Rasmussen}}, \bibinfo {author} {\bibfnamefont {H.~I.}\ \bibnamefont
  {J{\o}rgensen}}, \bibinfo {author} {\bibfnamefont {T.}~\bibnamefont
  {Hayashi}}, \bibinfo {author} {\bibfnamefont {P.~E.}\ \bibnamefont
  {Lindelof}}, \ and\ \bibinfo {author} {\bibfnamefont {T.}~\bibnamefont
  {Fujisawa}},\ }\href {\doibase 10.1021/nl072948y} {\bibfield  {journal}
  {\bibinfo  {journal} {Nano Lett.}\ }\textbf {\bibinfo {volume} {8}},\
  \bibinfo {pages} {1055} (\bibinfo {year} {2008})}\BibitemShut {NoStop}%
\bibitem [{\citenamefont {Kuzmenko}\ \emph {et~al.}(2006)\citenamefont
  {Kuzmenko}, \citenamefont {Kikoin},\ and\ \citenamefont
  {Avishai}}]{KuzmenkoPRL2006}%
  \BibitemOpen
  \bibfield  {author} {\bibinfo {author} {\bibfnamefont {T.}~\bibnamefont
  {Kuzmenko}}, \bibinfo {author} {\bibfnamefont {K.}~\bibnamefont {Kikoin}}, \
  and\ \bibinfo {author} {\bibfnamefont {Y.}~\bibnamefont {Avishai}},\ }\href
  {\doibase 10.1103/PhysRevLett.96.046601} {\bibfield  {journal} {\bibinfo
  {journal} {Phys. Rev. Lett.}\ }\textbf {\bibinfo {volume} {96}},\ \bibinfo
  {pages} {046601} (\bibinfo {year} {2006})}\BibitemShut {NoStop}%
\bibitem [{\citenamefont {Trocha}\ and\ \citenamefont
  {Barna\ifmmode~\acute{s}\else \'{s}\fi{}}(2008)}]{TrochaPRB2008}%
  \BibitemOpen
  \bibfield  {author} {\bibinfo {author} {\bibfnamefont {P.}~\bibnamefont
  {Trocha}}\ and\ \bibinfo {author} {\bibfnamefont {J.}~\bibnamefont
  {Barna\ifmmode~\acute{s}\else \'{s}\fi{}}},\ }\href {\doibase
  10.1103/PhysRevB.78.075424} {\bibfield  {journal} {\bibinfo  {journal} {Phys.
  Rev. B}\ }\textbf {\bibinfo {volume} {78}},\ \bibinfo {pages} {075424}
  (\bibinfo {year} {2008})}\BibitemShut {NoStop}%
\bibitem [{\citenamefont {Rogge}\ and\ \citenamefont
  {Haug}(2008)}]{RoggePRB2008}%
  \BibitemOpen
  \bibfield  {author} {\bibinfo {author} {\bibfnamefont {M.~C.}\ \bibnamefont
  {Rogge}}\ and\ \bibinfo {author} {\bibfnamefont {R.~J.}\ \bibnamefont
  {Haug}},\ }\href {\doibase 10.1103/PhysRevB.77.193306} {\bibfield  {journal}
  {\bibinfo  {journal} {Phys. Rev. B}\ }\textbf {\bibinfo {volume} {77}},\
  \bibinfo {pages} {193306} (\bibinfo {year} {2008})}\BibitemShut {NoStop}%
\bibitem [{\citenamefont {Breuer}\ and\ \citenamefont
  {Petruccione}(2006)}]{BreuerBook2006}%
  \BibitemOpen
  \bibfield  {author} {\bibinfo {author} {\bibfnamefont {H.-P.}\ \bibnamefont
  {Breuer}}\ and\ \bibinfo {author} {\bibfnamefont {F.}~\bibnamefont
  {Petruccione}},\ }\href@noop {} {\emph {\bibinfo {title} {Open Quantum
  Systems}}}\ (\bibinfo  {publisher} {Oxford University Press},\ \bibinfo
  {address} {Oxford},\ \bibinfo {year} {2006})\BibitemShut {NoStop}%
\bibitem [{\citenamefont {Parson}(2007)}]{ParsonBook2007}%
  \BibitemOpen
  \bibfield  {author} {\bibinfo {author} {\bibfnamefont {W.~W.}\ \bibnamefont
  {Parson}},\ }\href@noop {} {\emph {\bibinfo {title} {Modern Optical
  Spectroscopy}}}\ (\bibinfo  {publisher} {Springer Berlin},\ \bibinfo {year}
  {2007})\BibitemShut {NoStop}%
\bibitem [{\citenamefont {Lindgren}\ \emph {et~al.}(2013)\citenamefont
  {Lindgren}, \citenamefont {Kawaguchi}, \citenamefont {Heurlin}, \citenamefont
  {Borgstr{\"o}m}, \citenamefont {Pistol}, \citenamefont {Samuelson},\ and\
  \citenamefont {Gustafsson}}]{LindgrenNT2013}%
  \BibitemOpen
  \bibfield  {author} {\bibinfo {author} {\bibfnamefont {D.}~\bibnamefont
  {Lindgren}}, \bibinfo {author} {\bibfnamefont {K.}~\bibnamefont {Kawaguchi}},
  \bibinfo {author} {\bibfnamefont {M.}~\bibnamefont {Heurlin}}, \bibinfo
  {author} {\bibfnamefont {M.~T.}\ \bibnamefont {Borgstr{\"o}m}}, \bibinfo
  {author} {\bibfnamefont {M.-E.}\ \bibnamefont {Pistol}}, \bibinfo {author}
  {\bibfnamefont {L.}~\bibnamefont {Samuelson}}, \ and\ \bibinfo {author}
  {\bibfnamefont {A.}~\bibnamefont {Gustafsson}},\ }\href {\doibase
  10.1088/0957-4484/24/22/225203} {\bibfield  {journal} {\bibinfo  {journal}
  {Nanotechnology}\ }\textbf {\bibinfo {volume} {24}},\ \bibinfo {pages}
  {225203} (\bibinfo {year} {2013})}\BibitemShut {NoStop}%
\bibitem [{\citenamefont {Jurgilaitis}\ \emph {et~al.}(2014)\citenamefont
  {Jurgilaitis}, \citenamefont {Enquist}, \citenamefont {Harb}, \citenamefont
  {Dick}, \citenamefont {Borg}, \citenamefont {N{\"u}ske}, \citenamefont
  {Wernersson},\ and\ \citenamefont
  {Larsson}}]{JurgilaitisStructuralDynamics2014}%
  \BibitemOpen
  \bibfield  {author} {\bibinfo {author} {\bibfnamefont {A.}~\bibnamefont
  {Jurgilaitis}}, \bibinfo {author} {\bibfnamefont {H.}~\bibnamefont
  {Enquist}}, \bibinfo {author} {\bibfnamefont {M.}~\bibnamefont {Harb}},
  \bibinfo {author} {\bibfnamefont {K.~A.}\ \bibnamefont {Dick}}, \bibinfo
  {author} {\bibfnamefont {B.~M.}\ \bibnamefont {Borg}}, \bibinfo {author}
  {\bibfnamefont {R.}~\bibnamefont {N{\"u}ske}}, \bibinfo {author}
  {\bibfnamefont {L.-E.}\ \bibnamefont {Wernersson}}, \ and\ \bibinfo {author}
  {\bibfnamefont {J.}~\bibnamefont {Larsson}},\ }\href {\doibase
  10.1063/1.4833559} {\bibfield  {journal} {\bibinfo  {journal} {J. Struct.
  Dyn.}\ }\textbf {\bibinfo {volume} {1}},\ \bibinfo {pages} {014502} (\bibinfo
  {year} {2014})}\BibitemShut {NoStop}%
\bibitem [{\citenamefont {Wacker}(2002{\natexlab{a}})}]{WackerPR2002}%
  \BibitemOpen
  \bibfield  {author} {\bibinfo {author} {\bibfnamefont {A.}~\bibnamefont
  {Wacker}},\ }\href {\doibase http://dx.doi.org/10.1016/S0370-1573(01)00029-1}
  {\bibfield  {journal} {\bibinfo  {journal} {Phys. Rep.}\ }\textbf {\bibinfo
  {volume} {357}},\ \bibinfo {pages} {1 } (\bibinfo {year}
  {2002}{\natexlab{a}})}\BibitemShut {NoStop}%
\bibitem [{\citenamefont {Joyce}\ \emph {et~al.}(2013)\citenamefont {Joyce},
  \citenamefont {Docherty}, \citenamefont {Gao}, \citenamefont {Tan},
  \citenamefont {Jagadish}, \citenamefont {Lloyd-Hughes}, \citenamefont
  {Herz},\ and\ \citenamefont {Johnston}}]{JoyceNanotechnology2013}%
  \BibitemOpen
  \bibfield  {author} {\bibinfo {author} {\bibfnamefont {H.~J.}\ \bibnamefont
  {Joyce}}, \bibinfo {author} {\bibfnamefont {C.~J.}\ \bibnamefont {Docherty}},
  \bibinfo {author} {\bibfnamefont {Q.}~\bibnamefont {Gao}}, \bibinfo {author}
  {\bibfnamefont {H.~H.}\ \bibnamefont {Tan}}, \bibinfo {author} {\bibfnamefont
  {C.}~\bibnamefont {Jagadish}}, \bibinfo {author} {\bibfnamefont
  {J.}~\bibnamefont {Lloyd-Hughes}}, \bibinfo {author} {\bibfnamefont {L.~M.}\
  \bibnamefont {Herz}}, \ and\ \bibinfo {author} {\bibfnamefont {M.~B.}\
  \bibnamefont {Johnston}},\ }\href@noop {} {\bibfield  {journal} {\bibinfo
  {journal} {Nanotechnology}\ }\textbf {\bibinfo {volume} {24}},\ \bibinfo
  {pages} {214006} (\bibinfo {year} {2013})}\BibitemShut {NoStop}%
\bibitem [{\citenamefont {Kinaret}\ \emph {et~al.}(1992)\citenamefont
  {Kinaret}, \citenamefont {Meir}, \citenamefont {Wingreen}, \citenamefont
  {Lee},\ and\ \citenamefont {Wen}}]{KinaretPRB1992}%
  \BibitemOpen
  \bibfield  {author} {\bibinfo {author} {\bibfnamefont {J.~M.}\ \bibnamefont
  {Kinaret}}, \bibinfo {author} {\bibfnamefont {Y.}~\bibnamefont {Meir}},
  \bibinfo {author} {\bibfnamefont {N.~S.}\ \bibnamefont {Wingreen}}, \bibinfo
  {author} {\bibfnamefont {P.~A.}\ \bibnamefont {Lee}}, \ and\ \bibinfo
  {author} {\bibfnamefont {X.-G.}\ \bibnamefont {Wen}},\ }\href {\doibase
  10.1103/PhysRevB.46.4681} {\bibfield  {journal} {\bibinfo  {journal}
  {Phys.~Rev.~B}\ }\textbf {\bibinfo {volume} {46}},\ \bibinfo {pages} {4681}
  (\bibinfo {year} {1992})}\BibitemShut {NoStop}%
\bibitem [{\citenamefont {Pfannkuche}\ and\ \citenamefont
  {Ulloa}(1995)}]{PfannkuchePRL1995}%
  \BibitemOpen
  \bibfield  {author} {\bibinfo {author} {\bibfnamefont {D.}~\bibnamefont
  {Pfannkuche}}\ and\ \bibinfo {author} {\bibfnamefont {S.~E.}\ \bibnamefont
  {Ulloa}},\ }\href {\doibase 10.1103/PhysRevLett.74.1194} {\bibfield
  {journal} {\bibinfo  {journal} {Phys.~Rev.~Lett.}\ }\textbf {\bibinfo
  {volume} {74}},\ \bibinfo {pages} {1194} (\bibinfo {year}
  {1995})}\BibitemShut {NoStop}%
\bibitem [{\citenamefont {Cavaliere}\ \emph {et~al.}(2009)\citenamefont
  {Cavaliere}, \citenamefont {Giovannini}, \citenamefont {Sassetti},\ and\
  \citenamefont {Kramer}}]{CavaliereNJP2009}%
  \BibitemOpen
  \bibfield  {author} {\bibinfo {author} {\bibfnamefont {F.}~\bibnamefont
  {Cavaliere}}, \bibinfo {author} {\bibfnamefont {U.~D.}\ \bibnamefont
  {Giovannini}}, \bibinfo {author} {\bibfnamefont {M.}~\bibnamefont
  {Sassetti}}, \ and\ \bibinfo {author} {\bibfnamefont {B.}~\bibnamefont
  {Kramer}},\ }\href {\doibase 10.1088/1367-2630/11/12/123004} {\bibfield
  {journal} {\bibinfo  {journal} {New Journal of Physics}\ }\textbf {\bibinfo
  {volume} {11}},\ \bibinfo {pages} {123004} (\bibinfo {year}
  {2009})}\BibitemShut {NoStop}%
\bibitem [{\citenamefont {Hewson}(1993)}]{HewsonBook1993}%
  \BibitemOpen
  \bibfield  {author} {\bibinfo {author} {\bibfnamefont {A.~C.}\ \bibnamefont
  {Hewson}},\ }\href@noop {} {\emph {\bibinfo {title} {{The Kondo Problem to
  Heavy Fermions}}}}\ (\bibinfo  {publisher} {Cambridge University Press},\
  \bibinfo {year} {1993})\BibitemShut {NoStop}%
\bibitem [{\citenamefont {Bockelmann}\ and\ \citenamefont
  {Bastard}(1990)}]{BockelmannPRB1990}%
  \BibitemOpen
  \bibfield  {author} {\bibinfo {author} {\bibfnamefont {U.}~\bibnamefont
  {Bockelmann}}\ and\ \bibinfo {author} {\bibfnamefont {G.}~\bibnamefont
  {Bastard}},\ }\href {\doibase 10.1103/PhysRevB.42.8947} {\bibfield  {journal}
  {\bibinfo  {journal} {Phys.~Rev.~B}\ }\textbf {\bibinfo {volume} {42}},\
  \bibinfo {pages} {8947} (\bibinfo {year} {1990})}\BibitemShut {NoStop}%
\bibitem [{\citenamefont {Agam}\ \emph {et~al.}(1997)\citenamefont {Agam},
  \citenamefont {Wingreen}, \citenamefont {Altshuler}, \citenamefont {Ralph},\
  and\ \citenamefont {Tinkham}}]{AgamPRL1997}%
  \BibitemOpen
  \bibfield  {author} {\bibinfo {author} {\bibfnamefont {O.}~\bibnamefont
  {Agam}}, \bibinfo {author} {\bibfnamefont {N.~S.}\ \bibnamefont {Wingreen}},
  \bibinfo {author} {\bibfnamefont {B.~L.}\ \bibnamefont {Altshuler}}, \bibinfo
  {author} {\bibfnamefont {D.~C.}\ \bibnamefont {Ralph}}, \ and\ \bibinfo
  {author} {\bibfnamefont {M.}~\bibnamefont {Tinkham}},\ }\href {\doibase
  10.1103/PhysRevLett.78.1956} {\bibfield  {journal} {\bibinfo  {journal}
  {Phys. Rev. Lett.}\ }\textbf {\bibinfo {volume} {78}},\ \bibinfo {pages}
  {1956} (\bibinfo {year} {1997})}\BibitemShut {NoStop}%
\bibitem [{\citenamefont {Nakaoka}\ \emph {et~al.}(2006)\citenamefont
  {Nakaoka}, \citenamefont {Clark}, \citenamefont {Krenner}, \citenamefont
  {Sabathil}, \citenamefont {Bichler}, \citenamefont {Arakawa}, \citenamefont
  {Abstreiter},\ and\ \citenamefont {Finley}}]{NakaokaPRB2006}%
  \BibitemOpen
  \bibfield  {author} {\bibinfo {author} {\bibfnamefont {T.}~\bibnamefont
  {Nakaoka}}, \bibinfo {author} {\bibfnamefont {E.~C.}\ \bibnamefont {Clark}},
  \bibinfo {author} {\bibfnamefont {H.~J.}\ \bibnamefont {Krenner}}, \bibinfo
  {author} {\bibfnamefont {M.}~\bibnamefont {Sabathil}}, \bibinfo {author}
  {\bibfnamefont {M.}~\bibnamefont {Bichler}}, \bibinfo {author} {\bibfnamefont
  {Y.}~\bibnamefont {Arakawa}}, \bibinfo {author} {\bibfnamefont
  {G.}~\bibnamefont {Abstreiter}}, \ and\ \bibinfo {author} {\bibfnamefont
  {J.~J.}\ \bibnamefont {Finley}},\ }\href {\doibase
  10.1103/PhysRevB.74.121305} {\bibfield  {journal} {\bibinfo  {journal} {Phys.
  Rev. B}\ }\textbf {\bibinfo {volume} {74}},\ \bibinfo {pages} {121305}
  (\bibinfo {year} {2006})}\BibitemShut {NoStop}%
\bibitem [{\citenamefont {Lassen}\ and\ \citenamefont
  {Wacker}(2007)}]{LassenPRB2007}%
  \BibitemOpen
  \bibfield  {author} {\bibinfo {author} {\bibfnamefont {B.}~\bibnamefont
  {Lassen}}\ and\ \bibinfo {author} {\bibfnamefont {A.}~\bibnamefont
  {Wacker}},\ }\href {\doibase 10.1103/PhysRevB.76.075316} {\bibfield
  {journal} {\bibinfo  {journal} {Phys.~Rev.~B}\ }\textbf {\bibinfo {volume}
  {76}},\ \bibinfo {pages} {075316} (\bibinfo {year} {2007})}\BibitemShut
  {NoStop}%
\bibitem [{\citenamefont {Wacker}\ and\ \citenamefont
  {Jauho}(1998)}]{WackerPRL1998}%
  \BibitemOpen
  \bibfield  {author} {\bibinfo {author} {\bibfnamefont {A.}~\bibnamefont
  {Wacker}}\ and\ \bibinfo {author} {\bibfnamefont {A.-P.}\ \bibnamefont
  {Jauho}},\ }\href@noop {} {\bibfield  {journal} {\bibinfo  {journal}
  {Phys.~Rev.~Lett.}\ }\textbf {\bibinfo {volume} {80}},\ \bibinfo {pages}
  {369} (\bibinfo {year} {1998})}\BibitemShut {NoStop}%
\bibitem [{\citenamefont {Callebaut}\ and\ \citenamefont
  {Hu}(2005)}]{CallebautJAP2005}%
  \BibitemOpen
  \bibfield  {author} {\bibinfo {author} {\bibfnamefont {H.}~\bibnamefont
  {Callebaut}}\ and\ \bibinfo {author} {\bibfnamefont {Q.}~\bibnamefont {Hu}},\
  }\href {\doibase http://dx.doi.org/10.1063/1.2136420} {\bibfield  {journal}
  {\bibinfo  {journal} {J.~Appl.~Phys.}\ }\textbf {\bibinfo {volume} {98}},\
  \bibinfo {pages} {104505} (\bibinfo {year} {2005})}\BibitemShut {NoStop}%
\bibitem [{\citenamefont {Bethe}(1947)}]{BethePR1947}%
  \BibitemOpen
  \bibfield  {author} {\bibinfo {author} {\bibfnamefont {H.~A.}\ \bibnamefont
  {Bethe}},\ }\href {\doibase 10.1103/PhysRev.72.339} {\bibfield  {journal}
  {\bibinfo  {journal} {Phys. Rev.}\ }\textbf {\bibinfo {volume} {72}},\
  \bibinfo {pages} {339} (\bibinfo {year} {1947})}\BibitemShut {NoStop}%
\bibitem [{\citenamefont {L\"{o}wdin}(1951)}]{LowdinJCP1951}%
  \BibitemOpen
  \bibfield  {author} {\bibinfo {author} {\bibfnamefont {P.-O.}\ \bibnamefont
  {L\"{o}wdin}},\ }\href {\doibase 10.1063/1.1748067} {\bibfield  {journal}
  {\bibinfo  {journal} {J. Chem. Phys.}\ }\textbf {\bibinfo {volume} {19}},\
  \bibinfo {pages} {1396} (\bibinfo {year} {1951})}\BibitemShut {NoStop}%
\bibitem [{\citenamefont {Fetter}\ and\ \citenamefont
  {Walecka}(2003)}]{FetterBook2003}%
  \BibitemOpen
  \bibfield  {author} {\bibinfo {author} {\bibfnamefont {A.~L.}\ \bibnamefont
  {Fetter}}\ and\ \bibinfo {author} {\bibfnamefont {J.~D.}\ \bibnamefont
  {Walecka}},\ }\href@noop {} {\emph {\bibinfo {title} {Quantum Theory of
  Many-particle Systems}}}\ (\bibinfo  {publisher} {Dover Publications},\
  \bibinfo {year} {2003})\BibitemShut {NoStop}%
\bibitem [{\citenamefont {Kir\v{s}anskas}\ \emph {et~al.}(2012)\citenamefont
  {Kir\v{s}anskas}, \citenamefont {Paaske},\ and\ \citenamefont
  {Flensberg}}]{KirsanskasPRB2012}%
  \BibitemOpen
  \bibfield  {author} {\bibinfo {author} {\bibfnamefont {G.}~\bibnamefont
  {Kir\v{s}anskas}}, \bibinfo {author} {\bibfnamefont {J.}~\bibnamefont
  {Paaske}}, \ and\ \bibinfo {author} {\bibfnamefont {K.}~\bibnamefont
  {Flensberg}},\ }\href {\doibase 10.1103/PhysRevB.86.075452} {\bibfield
  {journal} {\bibinfo  {journal} {Phys. Rev. B}\ }\textbf {\bibinfo {volume}
  {86}},\ \bibinfo {pages} {075452} (\bibinfo {year} {2012})}\BibitemShut
  {NoStop}%
\bibitem [{\citenamefont {Koller}\ \emph {et~al.}(2012)\citenamefont {Koller},
  \citenamefont {Grifoni},\ and\ \citenamefont {Paaske}}]{KollerPRB2012}%
  \BibitemOpen
  \bibfield  {author} {\bibinfo {author} {\bibfnamefont {S.}~\bibnamefont
  {Koller}}, \bibinfo {author} {\bibfnamefont {M.}~\bibnamefont {Grifoni}}, \
  and\ \bibinfo {author} {\bibfnamefont {J.}~\bibnamefont {Paaske}},\ }\href
  {\doibase 10.1103/PhysRevB.85.045313} {\bibfield  {journal} {\bibinfo
  {journal} {Phys. Rev. B}\ }\textbf {\bibinfo {volume} {85}},\ \bibinfo
  {pages} {045313} (\bibinfo {year} {2012})}\BibitemShut {NoStop}%
\bibitem [{\citenamefont {Pedersen}\ and\ \citenamefont
  {Wacker}(2005)}]{PedersenPRB2005a}%
  \BibitemOpen
  \bibfield  {author} {\bibinfo {author} {\bibfnamefont {J.~N.}\ \bibnamefont
  {Pedersen}}\ and\ \bibinfo {author} {\bibfnamefont {A.}~\bibnamefont
  {Wacker}},\ }\href {\doibase 10.1103/PhysRevB.72.195330} {\bibfield
  {journal} {\bibinfo  {journal} {Phys. Rev. B}\ }\textbf {\bibinfo {volume}
  {72}},\ \bibinfo {pages} {195330} (\bibinfo {year} {2005})}\BibitemShut
  {NoStop}%
\bibitem [{\citenamefont {Pedersen}\ and\ \citenamefont
  {Wacker}(2010)}]{PedersenPHE2010}%
  \BibitemOpen
  \bibfield  {author} {\bibinfo {author} {\bibfnamefont {J.~N.}\ \bibnamefont
  {Pedersen}}\ and\ \bibinfo {author} {\bibfnamefont {A.}~\bibnamefont
  {Wacker}},\ }\href@noop {} {\bibfield  {journal} {\bibinfo  {journal}
  {Physica~E}\ }\textbf {\bibinfo {volume} {42}},\ \bibinfo {pages} {595}
  (\bibinfo {year} {2010})}\BibitemShut {NoStop}%
\bibitem [{\citenamefont {Anders}(2008)}]{AndersPRL2008}%
  \BibitemOpen
  \bibfield  {author} {\bibinfo {author} {\bibfnamefont {F.~B.}\ \bibnamefont
  {Anders}},\ }\href {\doibase 10.1103/PhysRevLett.101.066804} {\bibfield
  {journal} {\bibinfo  {journal} {Phys. Rev. Lett.}\ }\textbf {\bibinfo
  {volume} {101}},\ \bibinfo {pages} {066804} (\bibinfo {year}
  {2008})}\BibitemShut {NoStop}%
\bibitem [{\citenamefont {Han}\ and\ \citenamefont {Heary}(2007)}]{HanPRL2007}%
  \BibitemOpen
  \bibfield  {author} {\bibinfo {author} {\bibfnamefont {J.~E.}\ \bibnamefont
  {Han}}\ and\ \bibinfo {author} {\bibfnamefont {R.~J.}\ \bibnamefont
  {Heary}},\ }\href {\doibase 10.1103/PhysRevLett.99.236808} {\bibfield
  {journal} {\bibinfo  {journal} {Phys. Rev. Lett.}\ }\textbf {\bibinfo
  {volume} {99}},\ \bibinfo {pages} {236808} (\bibinfo {year}
  {2007})}\BibitemShut {NoStop}%
\bibitem [{\citenamefont {Hershfield}(1993)}]{HershfieldPRL1993}%
  \BibitemOpen
  \bibfield  {author} {\bibinfo {author} {\bibfnamefont {S.}~\bibnamefont
  {Hershfield}},\ }\href {\doibase 10.1103/PhysRevLett.70.2134} {\bibfield
  {journal} {\bibinfo  {journal} {Phys. Rev. Lett.}\ }\textbf {\bibinfo
  {volume} {70}},\ \bibinfo {pages} {2134} (\bibinfo {year}
  {1993})}\BibitemShut {NoStop}%
\bibitem [{\citenamefont {Jin}\ \emph {et~al.}(2008)\citenamefont {Jin},
  \citenamefont {Zheng},\ and\ \citenamefont {Yan}}]{JinJCP2008}%
  \BibitemOpen
  \bibfield  {author} {\bibinfo {author} {\bibfnamefont {J.}~\bibnamefont
  {Jin}}, \bibinfo {author} {\bibfnamefont {X.}~\bibnamefont {Zheng}}, \ and\
  \bibinfo {author} {\bibfnamefont {Y.}~\bibnamefont {Yan}},\ }\href {\doibase
  10.1063/1.2938087} {\bibfield  {journal} {\bibinfo  {journal}
  {J.~Chem.~Phys.}\ }\textbf {\bibinfo {volume} {128}},\ \bibinfo {pages}
  {234703} (\bibinfo {year} {2008})}\BibitemShut {NoStop}%
\bibitem [{\citenamefont {Schoeller}(2009)}]{SchoellerEurPhysJSpecTop2009}%
  \BibitemOpen
  \bibfield  {author} {\bibinfo {author} {\bibfnamefont {H.}~\bibnamefont
  {Schoeller}},\ }\href {\doibase 10.1140/epjst/e2009-00962-3} {\bibfield
  {journal} {\bibinfo  {journal} {Eur. Phys. J. Spec. Top.}\ }\textbf {\bibinfo
  {volume} {168}},\ \bibinfo {pages} {179} (\bibinfo {year}
  {2009})}\BibitemShut {NoStop}%
\bibitem [{\citenamefont {Haldane}(1978)}]{HaldanePRL1978}%
  \BibitemOpen
  \bibfield  {author} {\bibinfo {author} {\bibfnamefont {F.~D.~M.}\
  \bibnamefont {Haldane}},\ }\href {\doibase 10.1103/PhysRevLett.40.416}
  {\bibfield  {journal} {\bibinfo  {journal} {Phys. Rev. Lett.}\ }\textbf
  {\bibinfo {volume} {40}},\ \bibinfo {pages} {416} (\bibinfo {year}
  {1978})}\BibitemShut {NoStop}%
\bibitem [{\citenamefont {Dorda}\ \emph {et~al.}(2014)\citenamefont {Dorda},
  \citenamefont {Nuss}, \citenamefont {von~der Linden},\ and\ \citenamefont
  {Arrigoni}}]{DordaPRB2014}%
  \BibitemOpen
  \bibfield  {author} {\bibinfo {author} {\bibfnamefont {A.}~\bibnamefont
  {Dorda}}, \bibinfo {author} {\bibfnamefont {M.}~\bibnamefont {Nuss}},
  \bibinfo {author} {\bibfnamefont {W.}~\bibnamefont {von~der Linden}}, \ and\
  \bibinfo {author} {\bibfnamefont {E.}~\bibnamefont {Arrigoni}},\ }\href
  {\doibase 10.1103/PhysRevB.89.165105} {\bibfield  {journal} {\bibinfo
  {journal} {Phys. Rev. B}\ }\textbf {\bibinfo {volume} {89}},\ \bibinfo
  {pages} {165105} (\bibinfo {year} {2014})}\BibitemShut {NoStop}%
\bibitem [{\citenamefont {Chen}\ \emph {et~al.}(2014)\citenamefont {Chen},
  \citenamefont {Hansen},\ and\ \citenamefont {Franco}}]{ChenJPhysChemC2014}%
  \BibitemOpen
  \bibfield  {author} {\bibinfo {author} {\bibfnamefont {L.}~\bibnamefont
  {Chen}}, \bibinfo {author} {\bibfnamefont {T.}~\bibnamefont {Hansen}}, \ and\
  \bibinfo {author} {\bibfnamefont {I.}~\bibnamefont {Franco}},\ }\href
  {\doibase 10.1021/jp505771f} {\bibfield  {journal} {\bibinfo  {journal} {J.
  Phys. Chem. C}\ }\textbf {\bibinfo {volume} {118}},\ \bibinfo {pages} {20009}
  (\bibinfo {year} {2014})}\BibitemShut {NoStop}%
\bibitem [{\citenamefont {Wacker}(2002{\natexlab{b}})}]{WackerPhysRep2002}%
  \BibitemOpen
  \bibfield  {author} {\bibinfo {author} {\bibfnamefont {A.}~\bibnamefont
  {Wacker}},\ }\href@noop {} {\bibfield  {journal} {\bibinfo  {journal}
  {Phys.~Rep.}\ }\textbf {\bibinfo {volume} {357}},\ \bibinfo {pages} {1}
  (\bibinfo {year} {2002}{\natexlab{b}})}\BibitemShut {NoStop}%
\bibitem [{\citenamefont {Lindskog}\ \emph {et~al.}(2014)\citenamefont
  {Lindskog}, \citenamefont {Wolf}, \citenamefont {Trinite}, \citenamefont
  {Liverini}, \citenamefont {Faist}, \citenamefont {Maisons}, \citenamefont
  {Carras}, \citenamefont {Aidam}, \citenamefont {Ostendorf},\ and\
  \citenamefont {Wacker}}]{LindskogAPL2014}%
  \BibitemOpen
  \bibfield  {author} {\bibinfo {author} {\bibfnamefont {M.}~\bibnamefont
  {Lindskog}}, \bibinfo {author} {\bibfnamefont {J.~M.}\ \bibnamefont {Wolf}},
  \bibinfo {author} {\bibfnamefont {V.}~\bibnamefont {Trinite}}, \bibinfo
  {author} {\bibfnamefont {V.}~\bibnamefont {Liverini}}, \bibinfo {author}
  {\bibfnamefont {J.}~\bibnamefont {Faist}}, \bibinfo {author} {\bibfnamefont
  {G.}~\bibnamefont {Maisons}}, \bibinfo {author} {\bibfnamefont
  {M.}~\bibnamefont {Carras}}, \bibinfo {author} {\bibfnamefont
  {R.}~\bibnamefont {Aidam}}, \bibinfo {author} {\bibfnamefont
  {R.}~\bibnamefont {Ostendorf}}, \ and\ \bibinfo {author} {\bibfnamefont
  {A.}~\bibnamefont {Wacker}},\ }\href {\doibase
  http://dx.doi.org/10.1063/1.4895123} {\bibfield  {journal} {\bibinfo
  {journal} {Appl.~Phys.~Lett.}\ }\textbf {\bibinfo {volume} {105}},\ \bibinfo
  {eid} {103106} (\bibinfo {year} {2014})}\BibitemShut {NoStop}%
\bibitem [{\citenamefont {Dupont}\ \emph {et~al.}(2012)\citenamefont {Dupont},
  \citenamefont {Fathololoumi}, \citenamefont {Wasilewski}, \citenamefont
  {Aers}, \citenamefont {Laframboise}, \citenamefont {Lindskog}, \citenamefont
  {Razavipour}, \citenamefont {Wacker}, \citenamefont {Ban},\ and\
  \citenamefont {Liu}}]{DupontJAP2012}%
  \BibitemOpen
  \bibfield  {author} {\bibinfo {author} {\bibfnamefont {E.}~\bibnamefont
  {Dupont}}, \bibinfo {author} {\bibfnamefont {S.}~\bibnamefont
  {Fathololoumi}}, \bibinfo {author} {\bibfnamefont {Z.~R.}\ \bibnamefont
  {Wasilewski}}, \bibinfo {author} {\bibfnamefont {G.}~\bibnamefont {Aers}},
  \bibinfo {author} {\bibfnamefont {S.~R.}\ \bibnamefont {Laframboise}},
  \bibinfo {author} {\bibfnamefont {M.}~\bibnamefont {Lindskog}}, \bibinfo
  {author} {\bibfnamefont {S.~G.}\ \bibnamefont {Razavipour}}, \bibinfo
  {author} {\bibfnamefont {A.}~\bibnamefont {Wacker}}, \bibinfo {author}
  {\bibfnamefont {D.}~\bibnamefont {Ban}}, \ and\ \bibinfo {author}
  {\bibfnamefont {H.~C.}\ \bibnamefont {Liu}},\ }\href {\doibase
  http://dx.doi.org/10.1063/1.3702571} {\bibfield  {journal} {\bibinfo
  {journal} {J.~Appl.~Phys.}\ }\textbf {\bibinfo {volume} {111}},\ \bibinfo
  {eid} {073111} (\bibinfo {year} {2012})}\BibitemShut {NoStop}%
\end{thebibliography}

%

\end{document}